\documentclass[]{pasj02} 
\usepackage[switch,mathlines]{lineno}
\usepackage{amsmath}
\usepackage{breqn}

\font\manual=manfnt at 7pt \def\dbend{\hbox{\raise0.9ex\hbox{\manual\char127\hspace{0.6em}}}}

\providecommand{\e}[1]{\ensuremath{\times 10^{#1}}}

\newcounter{INTERNALionstage}

\def\gtsim{\mathrel{\hbox{\rlap{\hbox{\lower4pt\hbox{$\sim$}}}\hbox{$>$}}}}
\def\lesssim{\mathrel{\hbox{\rlap{\hbox{\lower4pt\hbox{$\sim$}}}\hbox{$<$}}}}

\def\A{{\rm\thinspace \AA}}

\def\g{{\rm\thinspace g}}

\def\K{{\rm\thinspace K}}

\def\m{{\rm\thinspace m}}

\def\s{{\rm\thinspace s}}

\def\W{{\rm\thinspace W}}

\def\fexxiv{\mbox{{\rm Fe~{\sc xxiv}}}}
\def\fexxv{\mbox{{\rm Fe~{\sc xxv}}}}
\def\fexxvi{\mbox{{\rm Fe~{\sc xxvi}}}}
\def\nixxvii{\mbox{{\rm Ni~{\sc xxvii}}}}
\def\h0{\mbox{{\rm H}$^0$}}
\DeclareMathAlphabet{\vib}{OML}{cmm}{m}{it}

\renewcommand{\thefootnote}{\fnsymbol{footnote}}

\jyear{2024}
\Received{}
\Accepted{}

\graphicspath{{./}{figures/}}

\begin{document}

\title{XRISM Reveals Complex Multi-Temperature Structures in the Abell 2029 Galaxy Cluster
}

\author{
Arnab~Sarkar\altaffilmark{1}\altemailmark\orcid{0000-0002-5222-1337} \email{arnabsar@mit.edu}, 
Eric~Miller\altaffilmark{1}\orcid{0000-0002-3031-2326},
Naomi~Ota\altaffilmark{2}\orcid{0000-0002-2784-3652},
Caroline~Kilbourne\altaffilmark{3} \orcid{0000-0001-9464-4103},
Brian~McNamara\altaffilmark{4},
Ming Sun\altaffilmark{5} \orcid{0000-0001-5880-0703},
Lorenzo~Lovisari\altaffilmark{6,7}\orcid{0000-0002-3754-2415},
Stefano~Ettori\altaffilmark{8,9}\orcid{0000-0003-4117-8617},
Dominique~Eckert\altaffilmark{10}\orcid{0000-0001-7917-3892},
Andrew Szymkowiak\altaffilmark{11} \orcid{0000-0002-4974-687X},
Tommaso~Bartalesi\altaffilmark{8, 12} \orcid{0009-0004-5838-2213},
and 
Michael {Loewenstein},\altaffilmark{13,14,15} \orcid{0000-0002-1661-4029}
}

\altaffiltext{1}{Kavli Institute for Astrophysics and Space Research,
Massachusetts Institute of Technology, 70 Vassar St, Cambridge, MA 02139}
\altaffiltext{2}{Department of Physics, Nara Women's University, Nara 630-8506, Japan}
\altaffiltext{3}{NASA / Goddard Space Flight Center, Greenbelt, MD 20771, USA}
\altaffiltext{4}{Department of Physics \& Astronomy, Waterloo Centre for Astrophysics, University of Waterloo, Ontario N2L 3G1, Canada}
\altaffiltext{5}{Department of Physics and Astronomy, The University of Alabama in Huntsville, Huntsville, AL 35899, USA}
\altaffiltext{6}{INAF, Istituto di Astrofisica Spaziale e Fisica Cosmica di Milano, via A. Corti 12, 20133 Milano, Italy}
\altaffiltext{7}{Center for Astrophysics $|$ Harvard $\&$ Smithsonian, 60 Garden Street, Cambridge, MA 02138, USA}
\altaffiltext{8}{INAF, Osservatorio di Astrofisica e Scienza dello Spazio, via Piero Gobetti 93/3, 40129 Bologna, Italy}
\altaffiltext{9}{INFN, Sezione di Bologna, viale Berti Pichat 6/2, 40127 Bologna, Italy}
\altaffiltext{10}{Department of Astronomy, University of Geneva, Ch. d'Ecogia 16, CH-1290 Versoix, Switzerland}
\altaffiltext{11}{Yale Center for Astronomy and Astrophysics, Yale University, CT 06520-8121, USA}
\altaffiltext{12}{Dipartimento di Fisica e Astronomia “Augusto Righi” – Alma Mater Studiorum – Università di Bologna, via Gobetti 93/2, I-40129 Bologna}
\altaffiltext{13}{Department of Astronomy, University of Maryland, College Park, MD 20742, USA}
\altaffiltext{14}{NASA / Goddard Space Flight Center, Greenbelt, MD 20771, USA}
\altaffiltext{15}{Center for Research and Exploration in Space Science and Technology, NASA / GSFC (CRESST II), Greenbelt, MD 20771, USA}



\KeyWords{Galaxy Clusters: Intra-cluster medium --- }  

\maketitle

\begin{abstract}
The Resolve micro-calorimeter 
onboard XRISM is set to 
significantly advance our 
understanding of the complex 
intracluster medium (ICM) 
in galaxy clusters. 
We present $\sim$500 ks
XRISM observations covering the
central and two northern regions
of Abell~2029 galaxy cluster. 
Resolve enables us to distinguish
multiple emission lines 
from hydrogen-like and
helium-like iron (Fe) ions.
This study focuses on the 
multi-temperature structure of
Abell~2029 using line-ratio
diagnostics.
Using single-temperature 
collisionally ionized equilibrium
model,
we measure average plasma
temperatures of
6.73 keV, 
7.61 keV, and 8.14 keV in the central, 
inner northern, and outer northern regions, 
respectively,
spanning a radial range up to
700 kpc.
To further investigate thermal structure, 
we derive excitation and ionization 
temperatures by comparing observed 
emission-line flux ratios with atomic database 
predictions.
Significant deviations from the
single-temperature CIE model in the
central and inner northern regions
indicate the presence of multi-phase gas.
The excitation and ionization temperatures 
range from 2.85 keV to 8.5 keV in
the central region, 4.3 keV to 9.8 keV 
in the inner northern region, and 8.3 keV
to 10.4 keV in the outer northern region.
These temperature distributions are 
largely consistent with the previously
observed temperature gradient
of A2029. 
However, Resolve detects two notably 
cooler components--3.42 keV in the central
region and $\sim$4.3 keV in the
inner northern region--likely associated
with displaced cool gas due to gas sloshing.
{ Additionally, we thermally resolve a 
2.85 keV gas component at the 
core of A2029--potentially a significant
development in our understanding
of gas cooling.}
We propose that
this cooler gas is a direct product of
ongoing cooling processes in A2029,
having already cooled to its present 
temperature.
If this temperature structure is stable 
and no heating mechanism is present,
this reservoir is likely to 
cool to even lower temperatures
and 
form stars.

\end{abstract}

\section{Introduction}

X-ray astronomy has entered a new
precision era with the advent of
XRISM \citep{2020SPIE11444E..22T}.
At the heart of XRISM 
is the Resolve instrument, 
a cutting-edge X-ray microcalorimeter 
that combines high spectral 
resolution with imaging capabilities 
\citep{2022SPIE12181E..1SI}. 
Resolve enables detailed
studies of the physical conditions 
within galaxy clusters, 
such as temperature, density, 
turbulence, and
chemical composition, with 
unprecedented accuracy. 
By providing insights into 
the complex processes governing 
galaxy cluster evolution, 
including cooling flows, 
mergers, and feedback from 
supermassive black holes,
Resolve is set to revolutionize
our understanding of the 
largest gravitationally bound 
structures in the universe.

In this series of papers
(e.g., \cite{A2029_eric,A2029_naomi}),
we analyze XRISM observations 
of the galaxy cluster 
Abell 2029 (hereafter A2029). 
A2029 is a massive cool-core
cluster located at a 
redshift of \( z = 0.0787 \) 
\citep{1989ApJS...70....1A,Sohnetal2019a,Sohnetal2019b}. 
It is classified as a Bautz-Morgan 
type I and richness class 4.4 
cluster \citep{1978ApJ...226...55D}. 
As one of the most extensively 
studied clusters in the universe, 
A2029 has been investigated across 
multiple wavelengths, including 
photometry, spectroscopy, and weak-
lensing observations 
(e.g., \cite{1991ApJ...369...46U,Clarke2004,PaternoMahleretal2013,2018ApJ...855..100S}).
Optical observations reveal
that its central cD galaxy, 
IC 1101, extends over 600 kpc, 
making it one of the largest known 
galaxies 
\citep{1991ApJ...369...46U}.
Radio observations indicate
that the central region of
A2029 hosts the steep-spectrum 
radio source PKS 1508+059, 
characterized by a C-shaped
wide-angle-tailed (WAT) 
morphology 
\citep{1994ApJS...95..345T, Clarke2004}. 
WAT sources are typically 
associated with the brightest
cluster galaxies in merging 
clusters 
\citep{2023ApJ...944..132S, 2023Galax..11...67O, 2023Galax..11...73B}.
Chandra X-ray observations have 
revealed a gas-sloshing spiral 
structure in A2029, further 
supporting the scenario of a minor
merger at $>$ 1 Gyr ago 
\citep{2004ApJ...616..178C, PaternoMahleretal2013}. 
Additionally, multi-wavelength 
studies by 
\citet{2019ApJ...871..129S} 
identified two subsystems, A2033 and 
{ a southern
infalling group (SIG)}, currently accreting onto A2029. 
These subsystems are expected to 
merge with A2029 in the next few 
gigayears.
The intricate dynamical history
and anticipated future interactions 
make A2029 an exceptionally 
interesting system for
understanding galaxy cluster 
evolution.

In this paper,
we focus on investigating 
the multi-temperature structures 
in the ICM of A2029
using observations from
XRISM/Resolve 
micro-calorimeter
and { XRISM/Xtend CCD camera}
\citep{2025arXiv250208030N}. 
Readers interested 
in additional Resolve-based 
measurements of A2029 are 
encouraged to consult
Paper I \citep{A2029_eric}
and Paper II \citep{A2029_naomi} 
for a comprehensive overview.

In X-ray emitting plasma, 
temperatures derived from the 
6--7 keV energy band are 
primarily influenced by the 
ionization state,
as indicated by the 
$\fexxv$/$\fexxvi$ line ratio. 
In contrast, temperatures 
inferred from the
continuum 
are largely governed 
by the thermal 
bremsstrahlung emission
\citep{1988xrec.book.....S}. 
Discrepancies between these 
temperature measurements can 
arise from various physical 
processes. 
For example, cluster mergers 
may drive the ICM into a 
non-equilibrium ionization state
\citep{2022ApJ...935L..23S}, 
altering the $\fexxv$/$\fexxvi$
line ratio compared to 
the equilibrium conditions
assumed in this analysis 
\citep{2010A&A...509A..29P, 2010PASJ...62..335A}. 
Additionally, merger events
can accelerate electrons 
to suprathermal velocities,
deviating from the 
Maxwellian velocity distribution 
typically used in modeling
the continuum emission 
\citep{2009A&A...496...25P,2024ApJ...962..161S}.
Such non-Maxwellian distributions 
also impact the $\fexxv$/$\fexxvi$
line ratio 
\citep{2009A&A...503..373K}. 
Moreover, relativistic electron 
populations generated by strong 
merger shocks can contribute 
an extra continuum component 
through inverse Compton scattering 
of cosmic microwave
background photons 
\citep{1988xrec.book.....S}.

This paper is structured as
follows: Section 
\ref{sec:observation}
details the procedures 
used to process Resolve and Xtend data, 
including spectral extraction
and generation of the 
associated response files. 
It also describes the 
methods employed for 
spectral fitting, 
particularly the modeling of
sky-background, 
non-X-ray background 
contributions,
and accounting for
{ point spread function (PSF)} 
mixing between regions. 
Section \ref{sec:physics_line} 
explains the theoretical
framework behind line-ratio 
diagnostics used to estimate 
temperatures from emission 
line ratios.
Section \ref{sec:results} 
presents the results derived 
from the analysis, 
while Section \ref{sec:origin} 
provides a detailed
discussion of these findings. 
Finally, Section \ref{sec:summary} 
concludes the paper with a 
summary of our key results.
Throughout this study,
we adopt a $\Lambda$CDM 
cosmology with
\( H_0 = 70 \ \, \text{km s}^{-1} \, \text{Mpc}^{-1} \),
\( \Omega_m = 0.3 \), 
and \( \Omega_\Lambda = 0.7 \).
At the redshift of Abell 2029 
(\( z = 0.0787 \)), 1 arcminute 
corresponds to 89 kpc.
All redshifts and velocities 
are corrected to the
Solar System barycenter. 
The proto-solar abundance 
table from \citet{Lodders09} 
is used throughout this work.
Unless otherwise noted, 
all uncertainties are 
reported at the
1\( \sigma \) (68\%) 
confidence level.

\section{Observation}\label{sec:observation}

\subsection{Resolve data reduction}\label{sec:rslv_data_reduction}
XRISM observed the central 
region of Abell 2029 through
two co-aligned observations.
The first observation occurred
on January 10, 2024, with an 
exposure time of 12.44 ks 
(OBSID 000149000), followed
by a second observation on 
January 13, 2024, with an
exposure time of 25.11 ks
(OBSID 000151000).
Additionally, one of the 
outer pointings, designated North 1,
was observed between 
January 10, 2024, and 
January 13, 2024, with an
exposure time of 106.03 ks 
(OBSID 000150000). 
Another outer pointing, designated 
North 2, was observed between 
July 27, 2024, and August 
4, 2024, with an exposure 
time of 366.15 ks 
(OBSID 300053010). 
A detailed observation log
is provided in Table 
\ref{tab:obs_log}.
{ For this paper, 
we use data from
both Resolve and Xtend instruments.} 
The Resolve data were 
reprocessed using 
Heasoft-6.34 software, 
with latest calibration files 
from { CalDB v20241115} and default
screening procedures 
as outlined in the XRISM 
team`s analyses of 
N132D \citep{XRISM2024_N132D}
and Abell 2029 
\citep{XRISM_A2029_paper1}. 
The cleaned exposure times 
after screening are also
summarized in Table 
\ref{tab:obs_log}.

\begin{table*}
\caption{XRISM Observation logs\label{tab:obs_log}}   
\begin{center}
\setlength{\tabcolsep}{6pt}
\begin{tabular}{cccccc}
Pointing & Observation ID & RA & Decl. & Observation Date & Exposure\\
& & (deg) & (deg) & & (ks)\\
\hline
\hline
Central & 000149000 & 227.7341 & 5.7451 & 2024-01-10--2024-01-10 & 12.44\\
Central & 000151000 & 227.7331 & 5.7450 & 2024-01-13--2024-01-13 & 25.11\\
North 1 & 000150000 & 227.7635 & 5.7850 & 2024-01-10--2024-01-13 & 106.03\\
North 2 & 300053010 & 227.7943 & 5.8245 & 2024-07-27--2024-08-04 & 366.15\\
\hline
\end{tabular}
\end{center}
\end{table*}

{For detailed data reduction 
methods for Resolve
we refer readers to  Paper I 
\citep{A2029_eric}. 
In summary, 
the Resolve energy-scale drift 
during observations is 
monitored by periodically 
illuminating the focal 
plane with an $^{55}$Fe 
source mounted on the filter 
wheel, 
typically during 
Earth occultation. 
After gain reconstruction, 
the flat-field-averaged 
energy scale uncertainty 
is $\pm 0.3$ eV in the
5.4 -- 9 keV energy range based on in-orbit calibration integrations using Cr and Cu emission lines from a modulated X-ray source as well as Mn from $^{55}$Fe (Eckart et al. in prep).
To get a sense of the goodness of the energy-scale correction for a particular observation, 
the dedicated calibration pixel, 
located outside of the aperture and 
continuously illuminated by
a collimated $^{55}$Fe source, 
was used.
Gain reconstruction for
this pixel was performed only
using data from the fiducial intervals used for the rest of the array, 
followed by fitting the
Mn K$\alpha$ 5.9 keV 
calibration line 
during times excluding the fiducial intervals, 
simulating the observation 
of a celestial source 
\citep{XRISM2024_N132D}.  After this analysis, we found the worst line offset, 0.12 eV, in OBSID 000149000, which would be added in quadrature to the 0.3 eV systematic uncertainty. 

Uncertainty in the instrumental line widths comes from a combination of measurement uncertainty, time-dependent noise, and broadening induced by the time-dependent, energy-scale correction.  The calibration-pixel FWHM widths ranged from from 4.45 to 4.73 eV in these observations.  The current in-orbit calibration has a line-spread uncertainty of $\sim0.15$ eV from 6--7 keV.  This error at 6 keV results in a systematic uncertainty in turbulent velocity of  $\pm6\,{\rm km\,s^{-1}}$for turbulence of 50 kms$^{-1}$ and $\pm2\,{\rm km\,s^{-1}}$for turbulence of 150 kms$^{-1}$.

}




Spectra were extracted from 
the full array for each observation, 
excluding pixel 27, 
which exhibited unexpected 
scale jumps not accounted 
for by the $^{55}$Fe fiducial cadence. { Pixel 11 also experiences such abrupt changes, but infrequently.  We identified such a jump in the energy scale of pixel 11 during OBSID 000150000 (North 1), affecting 5 ks of exposure, via comparison to the gain trend on the other pixels, after much of the analysis had been completed.   To investigate the impact of this interval, we compared full-field analysis of the North 1 data with and without this interval of time and found no differences exceeding the statistical errors.  }

Only high-resolution primary
(Hp or \texttt{ITYPE=0}) 
events were included
in the analysis, 
as they constitute over
99\% of the 2--10 keV events
for each 
observation. 
This excludes low-resolution
secondary 
(Ls or \texttt{ITYPE=4}) events,
which primarily result from
instrumental effects at these 
low count rates. 
For each observation,
a non-X-ray background (NXB) 
spectrum was generated 
using the {\tt rslnxbgen} tool.
The NXB was derived from
the Resolve night-Earth data archive, using the same screening as applied to the source data,
with weighting applied based 
on the distribution of 
geomagnetic cut-off rigidity
sampled during the respective 
observation. 

\begin{figure*}
    \begin{tabular}{cc}
 \includegraphics[width=0.5\textwidth]{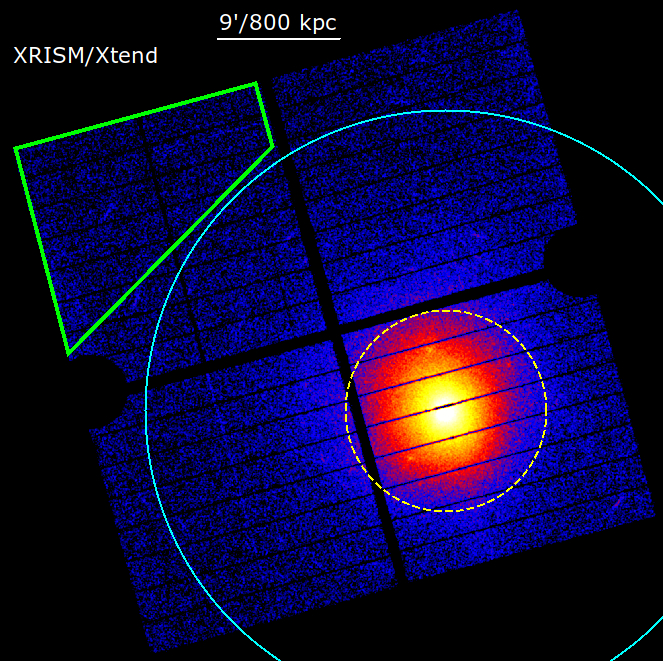} & \includegraphics[width=0.45\textwidth]{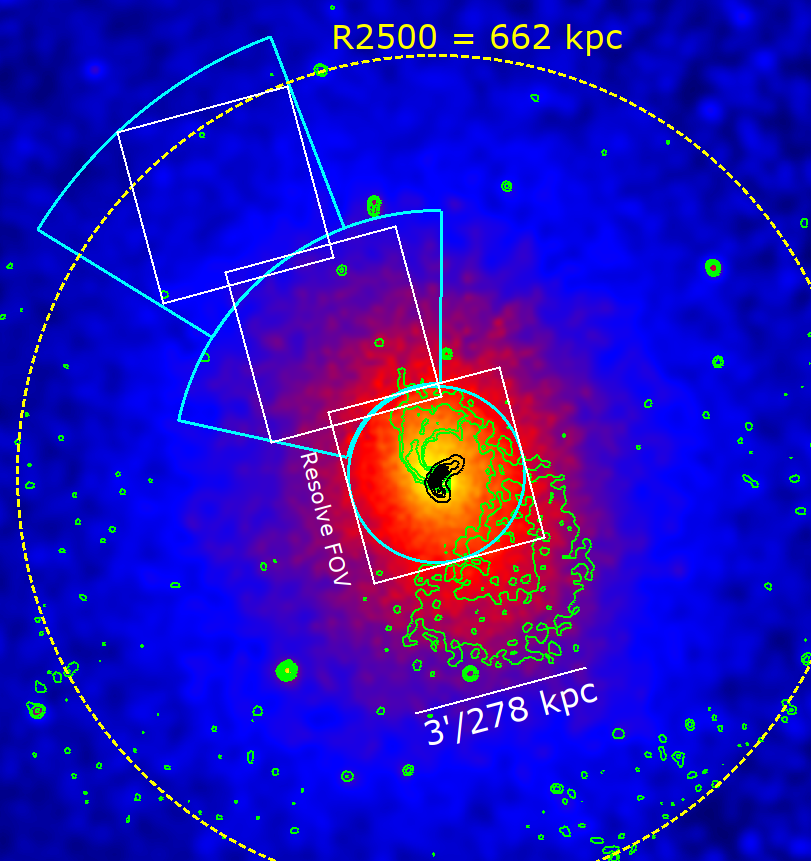}\\
  \end{tabular}
    \vspace{0.1in}
    \caption{Left: XRISM/Xtend 
    (Obs ID: 000150000) count 
    image of A2029 in the 
    0.5-10 keV energy band. 
    The yellow dashed and cyan 
    solid circles indicate
    R$_{2500}$ and R$_{200}$, 
    respectively. 
    The green rectangular region
    is used for the sky
    background analysis.
Right: Exposure-corrected,
background-subtracted Chandra
ACIS-I image of A2029.
White boxes represent the 
XRISM/Resolve field of views 
(FOVs) analyzed 
in this paper. 
Green surface brightness 
contours highlight the
cold-gas sloshing spiral 
in the cluster center,
while black contours show 
the 1.4 GHz radio emission
from the WAT 
{ active galactic nuclei (AGN)}
at 
the cluster core.
Cyan circle and sectors represent
the regions used for ancillary response files generation.
{Alt text: Left shows Xtend image of Abell 2029 and
right shows the Chandra image with three boxes arranged radially from the cluster core, indicating the positions of the Resolve's field of view.}}
\label{fig:chandra_image}
\end{figure*}

For each observations,
responses were generated using 
{\tt rslmkrmf} with the
latest CalDB. 
The { redistribution matrix file
(RMF)} was normalized by 
the Hp fraction in the
2--10 keV band, again 
ignoring Ls events, 
to account for the screened
medium- and low-resolution 
events and ensure proper flux 
normalization. 
The ancillary response 
files (ARF) were constructed
with {\tt xaarfgen},
using an exposure-corrected 
2--8 keV full-resolution 
Chandra image as the source model input,
as shown in Figure 
\ref{fig:chandra_image} (right).

\subsection{Xtend data reduction}\label{sec:xtnd_data_reduction}
{ For Xtend data reduction,
we followed the steps outlined 
in the quick start 
guide{\footnote{https://heasarc.gsfc.nasa.gov/docs/xrism/analysis/quickstart}}. 
Since in this paper we
used Xtend data to model the 
sky background, 
we only considered the 
North1 pointing (OBSID--000150000).
First, we identified and
filtered out the flickering
pixels from the event file. 
These high-count pixels,
caused by high-energy cosmic
rays and charged particles,
were removed by running 
{\tt searchflickpix} twice. 
This was necessary because
an updated version of
{\tt xtdflagpix} was required, 
and this latter tool had not
yet been released with
Heasoft 6.34.
We then masked the 
calibration sources on either
side of the Xtend FoV.
The flickering-pixel-removed
and calibration-source-masked 
event file was used to extract
an image in the 0.5--10 keV 
energy band, as shown in 
Figure 
\ref{fig:chandra_image} (left).
Spectrum was extracted from 
the rectangular region shown 
in 
Figure 
\ref{fig:chandra_image} (left).
Finally, we generated the 
response file using 
{\tt xtdrmf}, 
and a uniform ancillary 
response file 
was also generated using
{\tt xaarfgen}, applying
FLATCIRCLE mode with a
radius of $15\arcmin$.}

\subsection{Spectral analysis}\label{sec:spectral_analysis}
In this paper, we analyzed 
and fitted Resolve spectra 
extracted from the central
region and two northern 
pointings of Abell 2029. 
The detailed procedure for
spectral fitting is outlined
in the subsequent sections.

\subsubsection{Non-X-ray background}\label{sec:nxb_model}
The Non-X-ray Background (NXB) 
spectrum for each Resolve field of 
view (FOV) was extracted 
following the process described 
in Section 
\ref{sec:rslv_data_reduction}. 
The NXB spectrum
was modeled using a 
power-law component to represent
the continuum and several 
Gaussian components to account
for the instrumental fluorescence 
lines. 
During the spectral fitting 
of individual sources, 
the NXB model parameters 
were fixed to their 
best-fit values,
as detailed in Equations 
\ref{eq:spectra_central}--
\ref{eq:spectra_N2}.

For the Xtend instrument, 
the NXB spectrum
was similarly modeled. 
The continuum was represented
by a power-law with a fixed 
photon index of 0.25, 
while seven Gaussian components
were used to account for 
fluorescence lines with fixed
line centers at 2.15, 7.48, 9.71,
11.49, 11.48, and 13.41 keV.
For a comprehensive explanation
of the NXB extraction process 
and modeling approach, 
we refer readers to the
HEASARC 
webpage\footnote{https://heasarc.gsfc.nasa.gov/docs/xrism/analysis/nxb/}.

\subsubsection{Sky X-ray background}\label{sec:sky_bkg}
We modeled the sky background using 
XRISM Xtend observations of
A2029. 
With its large field of view, 
Xtend captures a significant 
portion of the sky, 
extending well beyond the 
$R_{200}$ radius of A2029 
(22.0$\arcmin$; 
\cite{Walkeretal2012a}), 
as illustrated in 
Figure \ref{fig:chandra_image} (left). 
To isolate the background,
we extracted source-free spectra
from the region highlighted 
in green in Figure 
\ref{fig:chandra_image} (left), 
located more than 
23 arcminutes from the
cluster center.
The response files 
were generated using 
the {\tt xtdrmf} tool, 
while a uniform ancillary 
response was created using
the {\tt xaarfgen} tool for a 
circular region with a radius
of 15 arcminutes. 
The X-ray emission from
this region is primarily
dominated by 
instrumental background, 
contributions from
Galactic foreground emission,
and cosmic X-ray background (CXB).

A2029 is positioned far from
the galactic plane,
resulting in a low hydrogen 
column density of $3 \times 10^{20}$
cm$^{-2}$
\citep{Walkeretal2012a}. 
This minimizes the impact 
of any column density 
fluctuations on soft X-ray 
emissions from the cluster. 
However, its proximity to the
North Polar Spur (NPS) 
introduces complications. 
The NPS, associated with 
hot interstellar bubbles
created by supernova 
explosions and stellar outflows, 
contributes significant 
soft X-ray emission
in this region
(e.g., \cite{2008PASJ...60S..95M,2024A&A...691L..22C,2024ApJ...967L..27L}). 
Consequently, previous 
studies have had to account
for the NPS's influence by
including additional 
background components to
model its contribution, 
alongside the typical
galactic emission background.

To address this, analyses 
of A2029 conducted with 
Chandra \citep{2005ApJ...628..655V} 
and XMM-Newton 
\citep{2008A&A...479..307B} 
employed a foreground 
emission model. 
This model featured two 
unabsorbed MEKAL components
with temperatures of
\(0.46 \pm 0.07 \, \text{keV}\) 
and \(0.22 \pm 0.05 \, \text{keV}\), 
providing a framework to
handle the complexities
introduced by the NPS.
Similarly, 
for Suzaku 
observations 
of same 
cluster,
\citet{Walkeretal2012a} found 
the galactic foreground
fits best with two unabsorbed
{\tt APEC}
models with $kT$ = 0.53 
$\pm$ 0.08 keV 
and 0.2 $\pm$ 0.05 keV. 
For this work, we closely follow
\citet{Walkeretal2012a} and
their treatment of sky background
since Xtend instrument has 
similar PSF with Suzaku.

To model the sky background, 
we fit the Xtend spectra using 
an absorbed power-law component 
with a fixed photon index of 
$\Gamma$ = 1.41 to represent 
the CXB \citep{2021MNRAS.501.3767S,2022MNRAS.516.3068S}.
Additionally, two unabsorbed 
{\tt APEC} models were 
included to account for
Galactic foreground emission, 
with temperatures fixed at 
$kT$ = 0.53 keV and $kT$ = 0.2 keV,
respectively 
\citep{Walkeretal2012a}. 
The NXB model was fixed based 
on the parameters outlined 
in Section \ref{sec:nxb_model}.

\subsubsection{Source spectrum modeling}
{ The Resolve instrument features
a modest point spread
function (PSF) of 
approximately {$1.3\arcmin$
\citep{2022SPIE12181E..1SI}}. 
During spectral fitting, 
it is crucial to account
for this PSF carefully, 
as X-ray photons from 
adjacent regions can 
substantially contaminate 
the spectrum of the target 
region. 
This effect, 
known as spatial-spectral 
mixing (SSM), 
must be considered to
ensure accurate spectral 
analysis.}
In this paper,
we explicitly address this
issue by accounting for 
SSM.
{ To accurately account
for SSM,
for each Resolve pointing, 
we generate three ancillary
response files (ARFs): 
one for that particular pointing 
and two for the rest of the 
pointings. 
The regions used for ARF 
extraction are illustrated 
in Figure \ref{fig:chandra_image}.
For example, for the
N2 pointing, 
we generate an ARF from 
the wedge-shaped region 
specific to N2. 
Additionally, we generate
two more ARFs—one from 
the N1 wedge and another
from the central circular 
region—since photons 
originating from the
N1 and central regions 
may contribute to the spectrum 
from N2 region. 
We calculate the fraction of 
detected X-ray photons that 
originate within the region
of interest versus those
originating from outside the
Resolve field of view, i.e., from
the other two pointings.
These calculations are
performed using 
ray-tracing simulations 
conducted with the
{\tt xrtraytrace} tool.}
Detailed results from the 
ray-tracing analysis are 
provided in Table~\ref{tab:ssm_fractio}.

\begin{table}
\caption{{ Raytracing results showing percentage of photon scattered from the regions in the columns into the regions in the rows.}\label{tab:ssm_fractio}}   
\begin{center}
\setlength{\tabcolsep}{6pt}
\begin{tabular}{cccc}
& Central & N1 & N2\\ 
\hline
Central & 74\% & 25\% & $<$1\%\\
N1 & 5\% & 63\% & 31\%\\
N2 & $<$1\% & 26\% & 73\%\\
\hline
\end{tabular}
\end{center}
\end{table}

Spectral fitting was performed
by using 
{\tt XSPEC v12.14.1} 
\citep{1996ASPC..101...17A} 
and by employing C-statistics 
\citep{1979ApJ...228..939C}. 
For plasma model calculations, 
we utilized the atomic databases 
from 
{\tt AtomDB v3.0.9} 
\citep{2012ApJ...756..128F} and 
{\tt SPEXACT v3.07.00} 
\citep{1996uxsa.conf..411K}. 
To ensure compatibility,
the {\tt SPEX} continuum and
line emission files were 
converted into a format 
readable by {\tt XSPEC}, 
allowing direct comparisons 
between the two databases 
while maintaining consistency
in the underlying assumptions
and fitting methodologies.
The spectrum extracted from 
each Resolve FOV was
fitted to a total of
five distinct components: 
(1) an absorbed velocity-broadened 
collisional equilibrium 
ionization (CIE)
model, {\tt TBabs$\times$bapec},
to model ICM emission from 
the specific region; 
(2) two additional absorbed 
{\tt bapec} models to account for
ICM emission originating from
the other two regions
contributing to the spectrum of
the region of interest;
(3) a sky-background model, as
described in Section 
\ref{sec:sky_bkg};
and (3) a power-law + Gaussian model 
to represent the NXB, as 
detailed in Section 
\ref{sec:nxb_model},

\begin{dmath}\label{eq:spectra_central}
    \rm Spectrum_{central} =  model_{central} \otimes ARF_{central}\ \\ +\ model_{N1} \otimes ARF_{N1 \rightarrow central}\ \\ +\ model_{N2} \otimes ARF_{N2 \rightarrow central}\ +\ NXB +\ BKG,
\end{dmath}

\begin{dmath}\label{eq:spectra_N1}
    \rm Spectrum_{N1} = model_{central} \otimes ARF_{central \rightarrow N1}\ \\ +\ model_{N1} \otimes ARF_{N1}\ \\ +\ model_{N2} \otimes ARF_{N2 \rightarrow N1}\ \\ +\ NXB +\ BKG,
\end{dmath}

\begin{dmath}\label{eq:spectra_N2}
    \rm Spectrum_{N2} = model_{central} \otimes ARF_{central \rightarrow N2}\ \\ +\ model_{N1} \otimes ARF_{N1 \rightarrow N2}\ \\ +\ model_{N2} \otimes ARF_{N2}\  \\ +\ NXB +\ BKG,
\end{dmath}
where ``model'' 
refers to the absorbed
{\tt bapec} model
for ICM emission and
the suffix `source 
$\rightarrow$ region' 
in each ARF indicates X-ray 
photons originating from the 
source but detected in the 
specified region. 
``BKG'' represents the sky 
background, as detailed in 
Section \ref{sec:sky_bkg}.
To account for the Galactic 
neutral column density along
the line of sight,
we applied the multiplicative 
{\tt TBabs} absorption model 
in {\tt XSPEC} 
\citep{2000ApJ...542..914W}, 
fixing $N_{\rm H}$ at 
$3\times10^{20}$ cm$^{-2}$ 
\citep{2016A&A...594A.116H}. 
Other parameters, 
including temperature, abundance, 
redshift, velocity broadening 
($\sigma_{\rm v}$), 
and normalization, 
were allowed to vary freely 
during the fitting.

\begin{figure*}
    \begin{tabular}{cc}
  \includegraphics[width=0.5\textwidth]{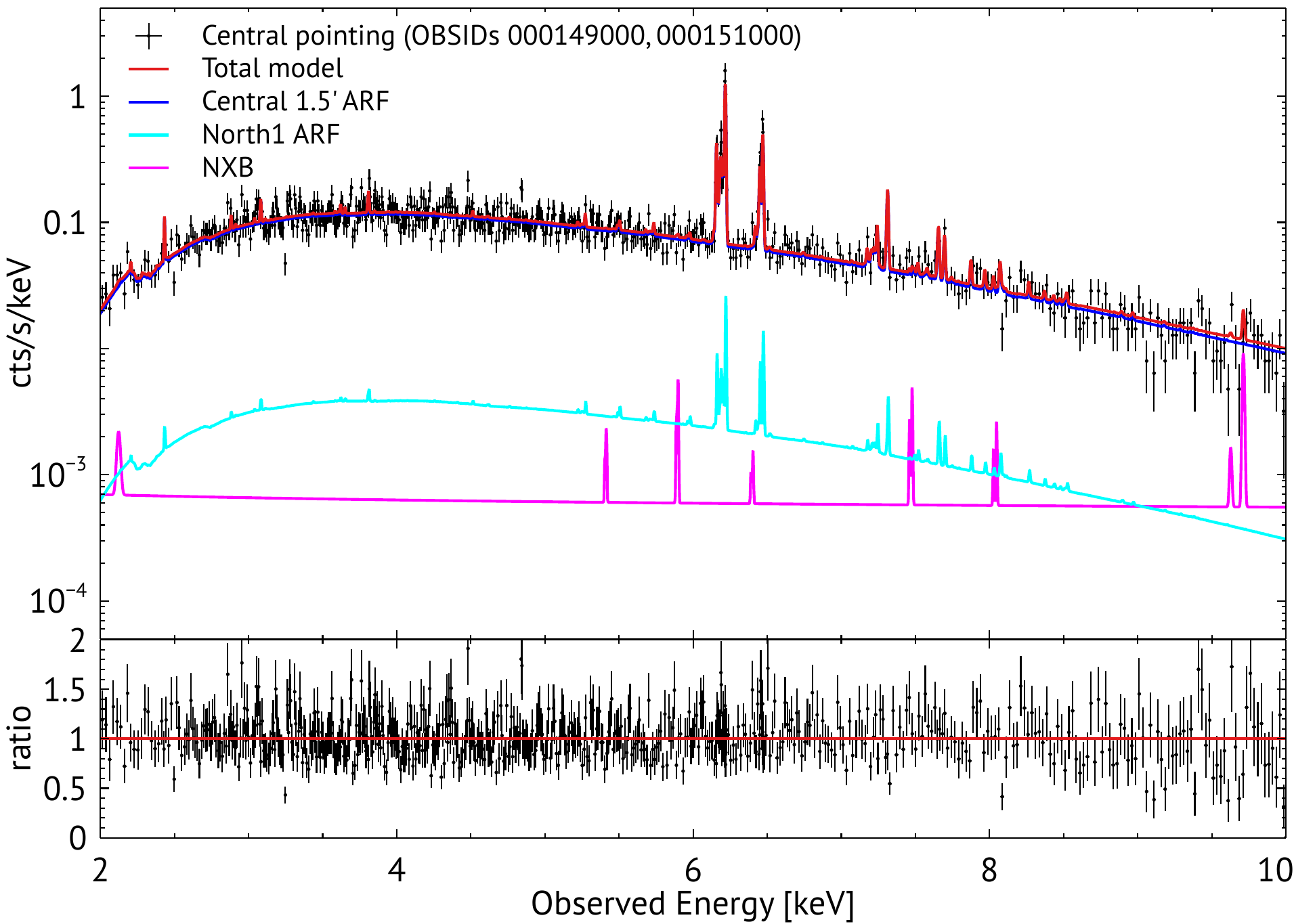} & \includegraphics[width=0.5\textwidth]{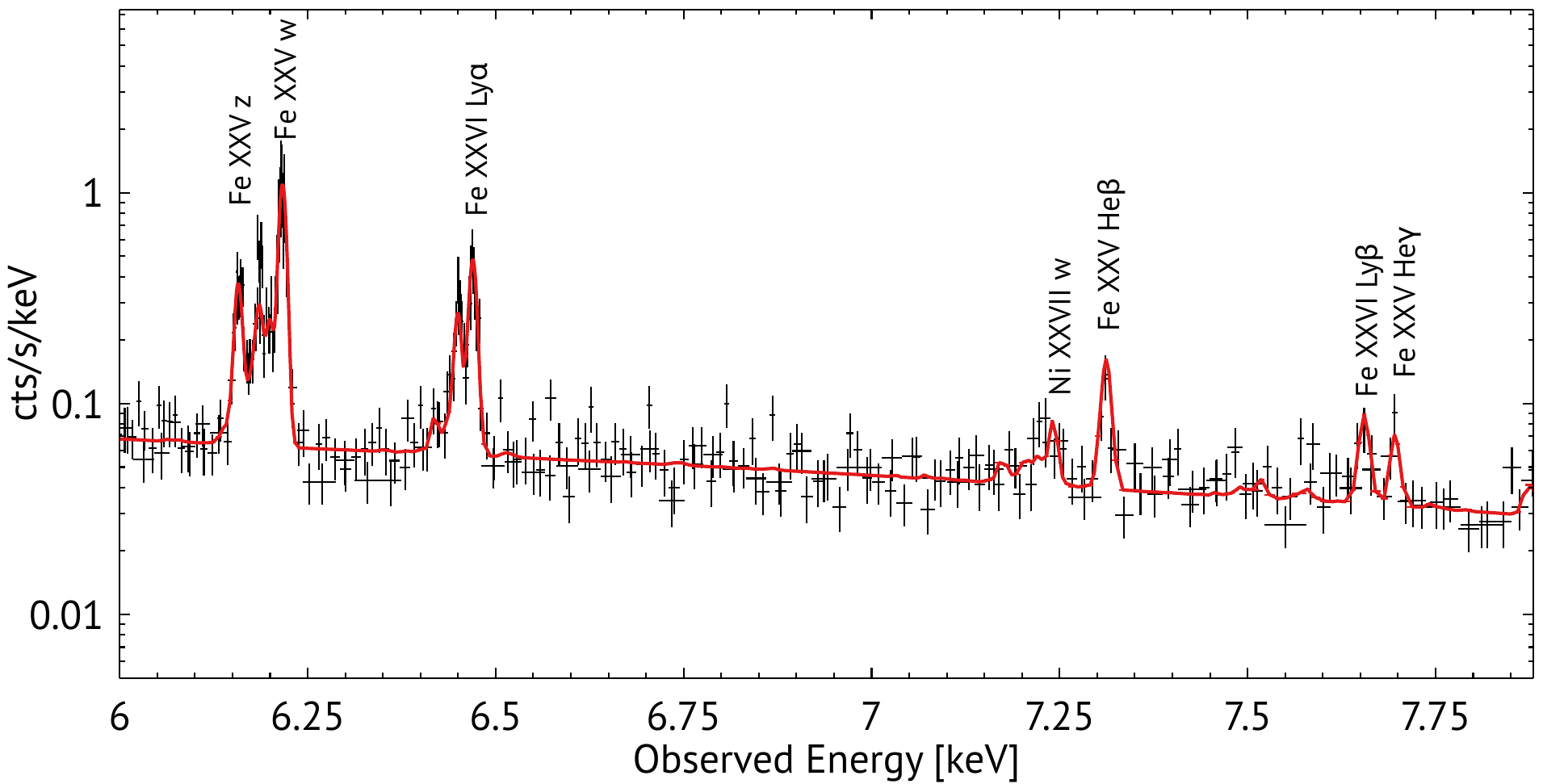}
  \end{tabular}
  \begin{tabular}{cc}
\includegraphics[width=0.5\textwidth]{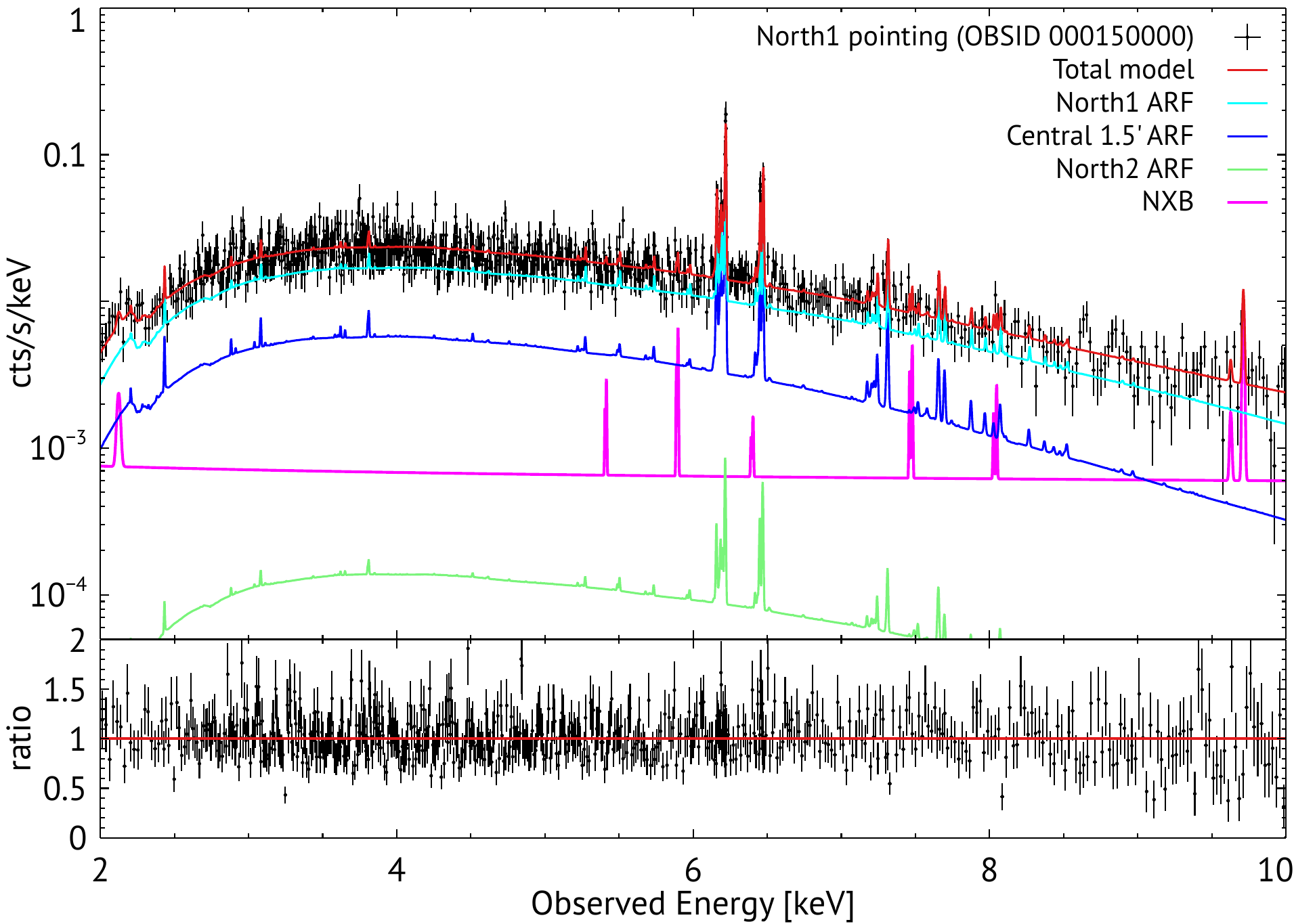} & \includegraphics[width=0.5\textwidth]{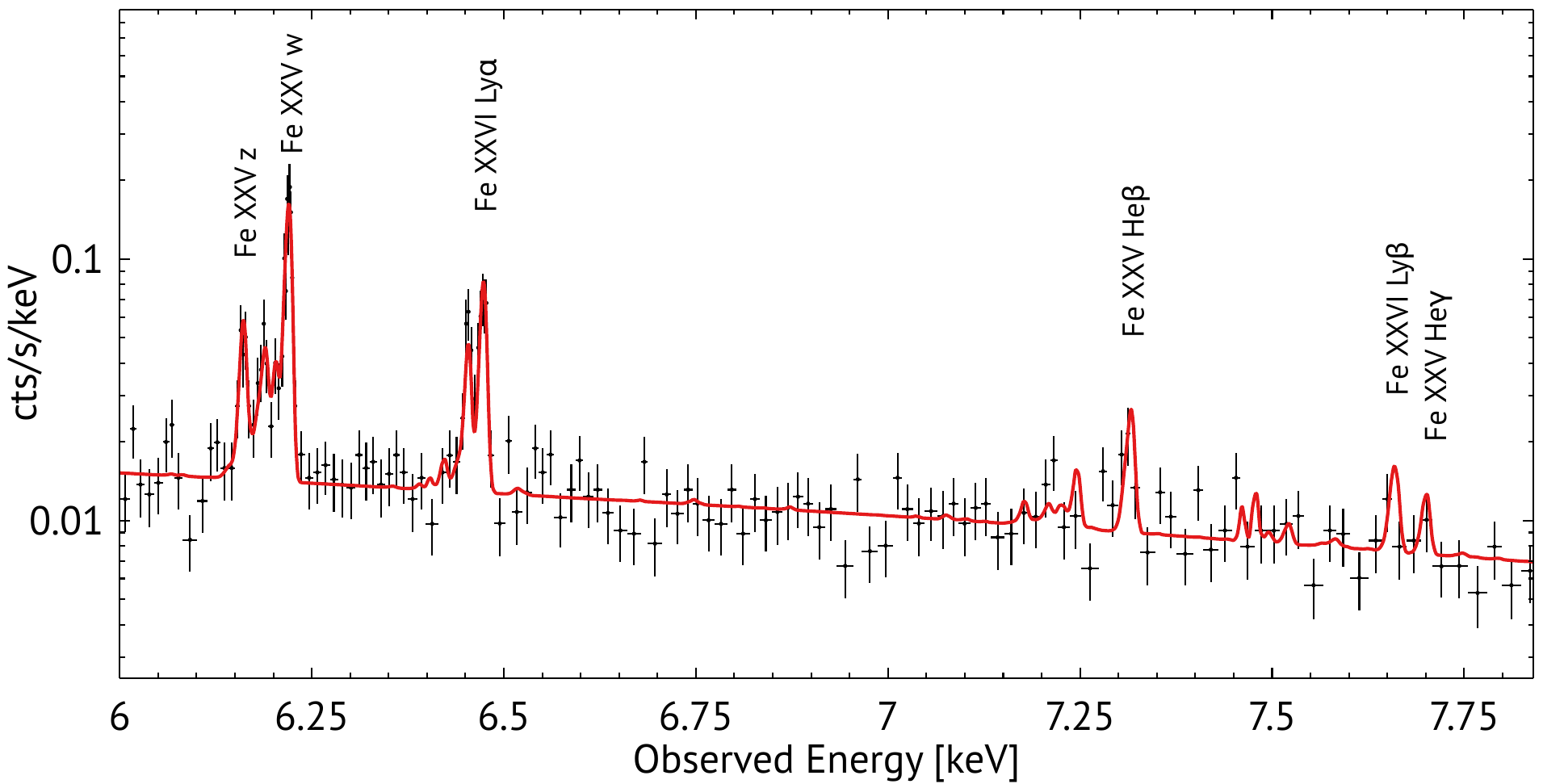}
  \end{tabular}
     \begin{tabular}{cc}
\includegraphics[width=0.5\textwidth]{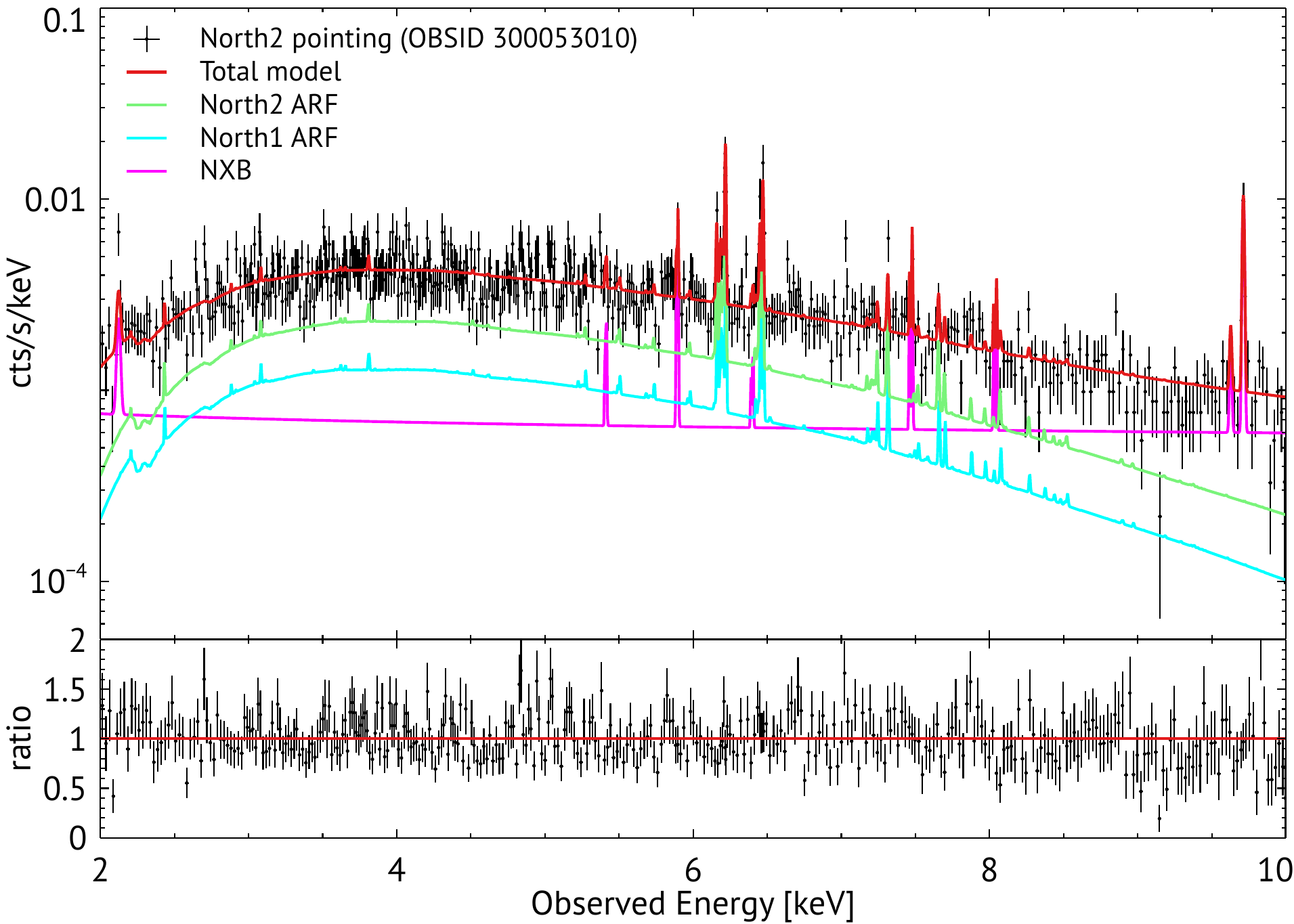} & \includegraphics[width=0.5\textwidth]{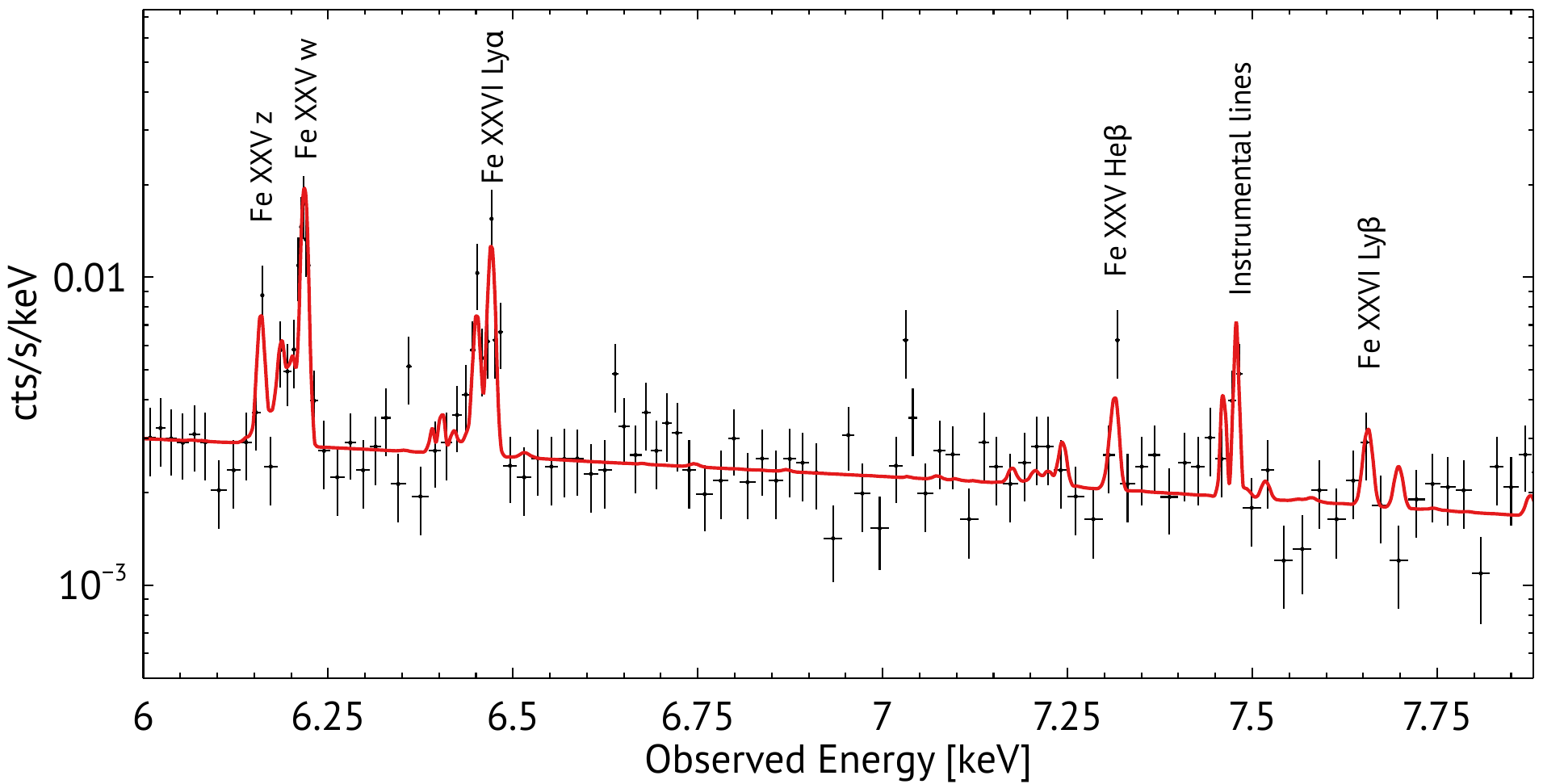}
    \end{tabular}
    \vspace{0.1in}
    \caption{{\it Left panels:} XRISM/Resolve spectra (black data points) from the central (top-left), N1 (middle-right), and N2 (bottom-left) regions, shown along with the best-fit models after accounting for Spatial-Spectral Mixing. In all three panels, the best-fit model components are represented as follows: red for the total model, blue for the central region, green for the N1 region, cyan for the N2 region, and magenta for the NXB. In all three panels, the bottom panels display the ratio between the observed data and the best-fit total models for the corresponding regions.
    {\it Right panels:} show
    a zoomed-in view of the left panels, focusing on the 6.5–8.4 keV (rest-frame) energy band, along with the best-fit total models.
    {Alt text: Left three vertical panels show Resolve spectra in 2--10 keV band and right three vertical panels show same but in 6--8 keV band.}
    }
    \label{fig:best_spec}
\end{figure*}

For the central pointing, 
all parameters were tied between
the spectra from both OBSIDs, 
as the observations were taken 
close enough in time that 
the barycentric corrections
($+25.5$ km,s$^{-1}$ for the 
first observation and 
$+26.0$ km,s$^{-1}$ for 
the second) were nearly identical. 
The spectral fitting was 
performed in the 2–10 keV 
energy band.
The best-fit spectra are 
displayed in Figure 
\ref{fig:best_spec},
and the corresponding 
best-fit ICM parameters are
provided in Table 
\ref{tab:best_params}.

\begin{table*}
\caption{Best-fit parameters from  Resolve spectral fitting 
including modeling
the Spatial-Spectral Mixing.\label{tab:best_params}}   
\begin{center}
\setlength{\tabcolsep}{6pt}
\begin{tabular}{lcllll}
Database & Model & Parameters & Central & N1 & N2\\
\hline
\hline
\\
& & kT (keV) & 6.73$^{+0.17}_{-0.15}$ & 7.61$^{+0.35}_{-0.33}$ & 8.14$^{+0.70}_{-0.65}$\\
& & Abundance ($Z_{\odot}$) & 0.64$^{+0.03}_{-0.03}$ & 0.34$^{+0.04}_{-0.04}$ & 0.41$^{+0.08}_{-0.07}$\\
{\tt AtomDB} & 1T & Redshift & 0.07776$^{+0.00004}_{-0.00003}$ & 0.07706$^{+0.00004}_{-0.00008}$ & 0.07769$^{+0.00012}_{-0.00012}$\\
& & $\sigma_{v}$ (km/s) & 157$^{+12}_{-13}$ & $<82$ & 92$^{+44}_{-44}$\\
& & norm (10$^{12}$ cm$^{-5}$) & 4.59$^{+0.08}_{-0.08}$ & 0.982$^{+0.03}_{-0.03}$ & 0.112$^{+0.006}_{-0.006}$\\
& & C-stat/dof & 16220/15999 & 15771/15999 & 15244/15999\\
\hline
\\
& & kT (keV) & 6.8$^{+0.17}_{-0.16}$ & 7.62$^{+0.34}_{-0.34}$ & 8.19$^{+0.78}_{-0.66}$\\
& & Abundance ($Z_{\odot}$) & 0.67$^{+0.04}_{-0.03}$ & 0.35$^{+0.03}_{-0.04}$ & 0.42$^{+0.08}_{-0.10}$\\
{\tt SPEXACT} & 1T & Redshift & 0.07777$^{+0.00004}_{-0.00004}$ & 0.07706$^{+0.00004}_{-0.00008}$ & 0.07768$^{+0.00011}_{-0.00011}$\\
& & $\sigma_{v}$ (km/s) & 157$^{+12}_{-13}$ & $<93$ & 92$^{+53}_{-53}$\\
& & norm (10$^{12}$ cm$^{-5}$) & 4.54$^{+0.09}_{-0.08}$ & 0.978$^{+0.030}_{-0.027}$ & 0.112$^{+0.007}_{-0.006}$\\
& & C-stat/dof & 16220/15999 & 15775/15999 & 15248/15999\\
\hline
\\
& & kT$_{1}$ (keV) & 5.14$^{+0.52}_{-1.30}$ & 3.60$^{+1.37}_{-1.23}$ & $-$\\
& & kT$_{2}$ (keV) & 8.66$^{+1.44}_{-1.01}$ & 8.70$^{+1.60}_{-0.92}$ & 8.14$^{+0.72}_{-0.54}$\\
& & Abundance ($Z_{\odot}$) & 0.68$^{+0.04}_{-0.03}$ & 0.36$^{+0.04}_{-0.04}$ & 0.41$^{+0.09}_{-0.07}$\\
{\tt AtomDB} & 2T & Redshift & 0.07776$^{+0.00004}_{-0.00003}$ & 0.07706$^{+0.00005}_{-0.00009}$ & 0.07769$^{+0.00012}_{-0.00012}$\\
& & $\sigma_{v}$ (km/s) & 155$^{+11}_{-11}$ & $<82$ & 94$^{+52}_{-52}$\\
& & norm$_{1}$ (10$^{12}$ cm$^{-5}$) & 2.29$^{+0.78}_{-0.72}$ & 0.186$^{+0.03}_{-0.02}$ & $-$\\
& & norm$_{2}$ (10$^{12}$ cm$^{-5}$) & 2.33$^{+0.81}_{-0.57}$ & 0.846$^{+0.03}_{-0.03}$ & 0.112$^{+0.007}_{-0.006}$\\
& & C-stat/dof & 16218/15999 & 15772/15999 & 15246/15999\\
\hline
\\
& & kT$_{1}$ (keV) & 4.15$^{+1.14}_{-1.31}$ & $-$ & $-$\\
& & kT$_{2}$ (keV) & 8.01$^{+1.87}_{-1.18}$ & 7.62$^{+0.35}_{-0.34}$ & 8.18$^{+0.78}_{-0.66}$\\
& & Abundance ($Z_{\odot}$) & 0.71$^{+0.05}_{-0.04}$ & 0.35$^{+0.04}_{-0.04}$ & 0.43$^{+0.10}_{-0.08}$\\
{\tt SPEXACT} & 2T & Redshift & 0.07777$^{+0.00004}_{-0.00003}$ & 0.07706$^{+0.00005}_{-0.00008}$ & 0.07770$^{+0.00011}_{-0.00011}$\\
& & $\sigma_{v}$ (km/s) & 156$^{+12}_{-13}$ & $<100$ & 93$^{+52}_{-52}$\\
& & norm$_{1}$ (10$^{12}$ cm$^{-5}$) & 1.42$^{+0.08}_{-0.08}$ & $-$ & $-$\\
& & norm$_{2}$ (10$^{12}$ cm$^{-5}$) & 3.24$^{+0.5}_{-0.5}$ & 0.978$^{+0.03}_{-0.02}$ & 0.111$^{+0.006}_{-0.006}$\\
& & C-stat/dof & 16219/15999 & 
15774/15999 & 15248/15999\\
\hline
\end{tabular}
\end{center}
\end{table*}

\section{Physics of line-ratio diagnostics}\label{sec:physics_line}
Spectral lines of H-like and 
He-like ions are among the 
most prominent features observed
in X-ray spectra across a wide 
range of astrophysical sources.
In CIE plasma, 
such as those found in 
galaxy clusters, 
the $\fexxv$ and $\fexxvi$ 
emission lines arise from 
transitions in He-like and 
H-like iron, respectively. 
These lines correspond to 
electron transitions from 
higher energy levels 
($n = 2, 3, 4, \dots$) to 
the ground state ($n = 1$).
In CIE plasma, 
electron-ion collisions serve
as the primary 
ionization mechanism, 
while electron impact 
predominantly populates 
the atomic energy levels. 
It is generally assumed 
that such plasmas are 
optically thin to their
own radiation (except resonance 
lines, such as $\fexxv$ w 
line) and that no 
external radiation field 
significantly influences the 
ionization balance 
\citep{2010SSRv..157..103P}.

The most prominent $\fexxv$ 
lines observed in galaxy clusters 
include the z, x, y, and w lines, 
collectively referred to as
the Fe He$\alpha$ complex 
\citep{2016Natur.535..117H, 2020ApJ...901...69C}. 
The x and y lines are 
intercombination lines, 
produced by magnetic quadrupole 
transitions corresponding to 
$1s2p\ (^3P_{2}) \rightarrow 1s^2\ 
(^1S_{0})$ 
and $1s2p\ (^3P_{1}) 
\rightarrow 1s^2\ (^1S_{0})$, 
respectively. 
The w line is a resonance line 
with the highest transition 
probability among the four,
arising from the electric dipole 
transition $1s2p\ (^1P_{1}) 
\rightarrow 1s^2\ (^1S_{0})$.
In contrast, the z line, 
a forbidden line, 
is temperature-sensitive and 
originates from the 
relativistic magnetic 
dipole transition 
$1s2s\ (^3S_{1}) 
\rightarrow 1s^2\ (^1S_{0})$.

In addition to the Fe 
He$\alpha$ complex, 
transitions involving higher 
principal quantum numbers 
(e.g., $n=3, 4, 5,\dots$) 
are also observed in the 
cores of galaxy clusters.
For instance, 
\citet{2016Natur.535..117H} 
reported the detection of Fe 
He$\beta$ ($1s3p\ (^1P_{1}) 
\rightarrow 1s^2\ (^1S_{0})$), 
Fe He$\gamma$ ($1s4p\ (^1P_{1}) 
\rightarrow 1s^2\ (^1S_{0})$), 
and Fe He$\epsilon$ ($1s5p\ (^1P_{1}) 
\rightarrow 1s^2\ (^1S_{0})$) lines in the Perseus cluster.
In Abell 2029, we detected the Fe 
He$\alpha$ complex and 
Fe He$\beta$ lines across
all three Resolve pointings. 
Additionally, we identified
the Fe He$\gamma$ line exclusively
in the central region with 
a statistical significance 
of $\gtrsim 3\sigma$, 
as illustrated in Figure 
\ref{fig:best_spec}.

Additionally, we detected
$\fexxvi$ lines (H-like Fe ions), 
specifically Fe Ly$\alpha$1, 
Fe Ly$\alpha$2, 
and Fe Ly$\beta$, 
in the spectra of all three 
regions in Abell 2029,
as shown in Figure 
\ref{fig:best_spec}.
In low-electron-density CIE plasmas, 
these lines are generally 
optically thin and are 
produced through collisional 
excitation followed by radiative 
transitions. 
The Fe Ly$\alpha$1 line 
originates from the
$2p_{3/2} \rightarrow 1s_{1/2}$ 
transition, 
the Fe Ly$\alpha$2 line from 
the $2p_{1/2} \rightarrow 1s_{1/2}$ 
transition, 
and the Fe Ly$\beta$ line from 
the $3p \rightarrow 1s_{1/2}$ 
transition.

Since these lines are
primarily produced by 
collisional excitations
of free electrons in the 
ICM plasma, 
the line-flux ratios between
specific pairs of $\fexxv$ 
and $\fexxvi$ lines can be 
used to measure the ICM plasma 
temperature. 
The excitation temperature, 
$T_{\rm exc}$, 
quantifies the energy of 
free electrons responsible 
for exciting bound electrons 
to higher energy levels and 
is directly related to the
line-flux ratios of similar 
ionization states 
(e.g., \cite{1991pav..book.....S,1985rpa..book.....R}).
\begin{equation}
    {\rm Ratio}_{u\rightarrow l} \propto  {\rm exp}(-\frac{\Delta E}{k_B T_{\rm exc}}),
\end{equation}
where 
$\Delta E$ is the energy difference
between the levels, $k_B$ is the
Boltzmann constant.
In a truly single-temperature
plasma,
the excitation temperature,
$T_{\rm exc}$, 
should be consistent with 
the electron temperature, $T_e$, 
derived from the bremsstrahlung 
continuum.
For Abell 2029, 
we calculate $T_{\rm exc}$
using the line ratios
$x+y+z/w$, He$\beta$/z, 
and He$\gamma$/z of Fe, 
where these lines are detected, 
as outlined in Section 
\ref{sec:temp_via_line}.

\begin{figure}
\includegraphics[width=0.45\textwidth]{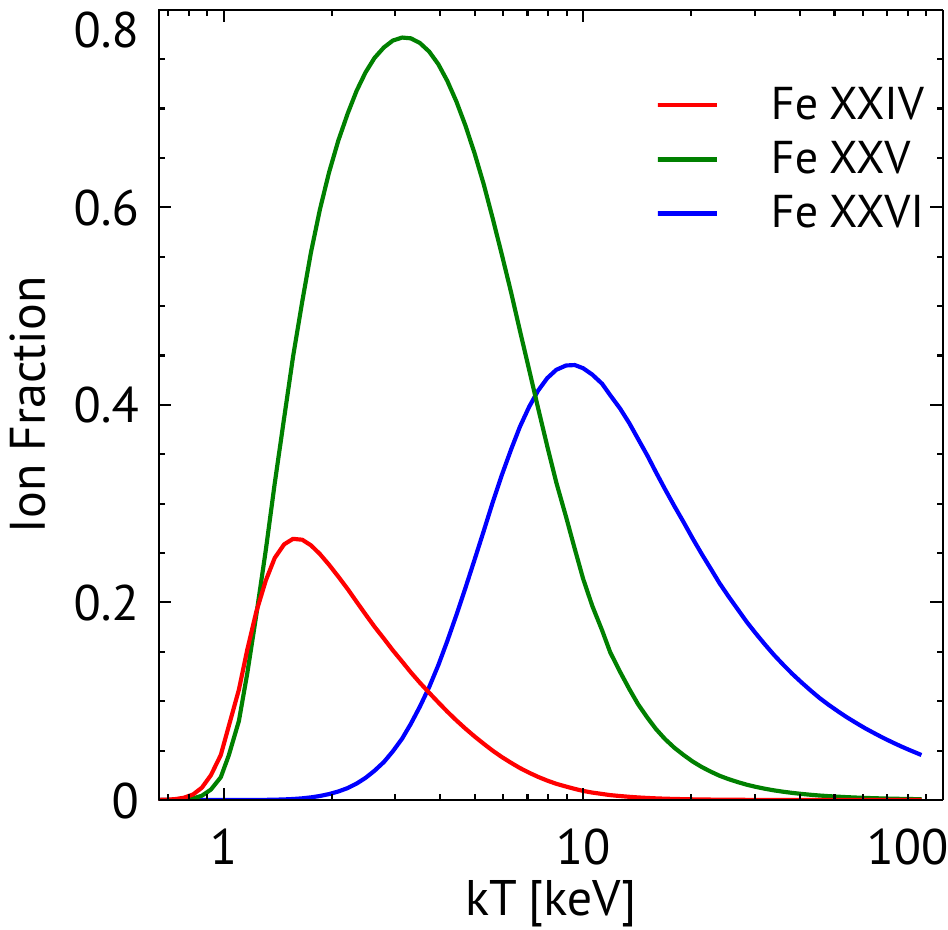}
    \caption{Ionization fractions
    of $\fexxiv$, $\fexxv$, 
    and $\fexxvi$ as a 
    function of temperature for
    a collisionally ionized plasma, calculated using \texttt{AtomDB}. Below $kT = 1$ keV ($10^7 \, \mathrm{K}$), the ionization fractions for all three ions are negligible, indicating that collisionally excited emission from these ions contributes minimally at lower temperatures.
    {Alt text: Three curves show ionization fractions of 
    $\fexxiv$, $\fexxv$, 
    and $\fexxvi$.}
    }
    \label{fig:ionization_vs_temp}
\end{figure}

The temperature dependence 
of the Fe ionization fraction 
provides an independent and 
reliable method for measuring
plasma temperature 
\citep{2010A&A...523A..22N, 2018PASJ...70...11H}. 
Figure \ref{fig:ionization_vs_temp} 
illustrates how the 
ionization fraction varies 
with plasma temperature for
H-like, He-like, and
Li-like Fe ions. 
The ionization temperature,
$T_{\rm ion}$, 
characterizes the relative 
population of different 
ionization states of an element. 
It is directly related to the 
line flux ratios between 
different ionization states
of the same element,
as described by the Saha equation
or its modifications for 
high-energy astrophysical plasmas
(e.g., \cite{1991pav..book.....S,1985rpa..book.....R}),
\begin{equation}
    {\rm Ratio}_{u\rightarrow l} \propto T_{\rm ion}^{3/2} {\rm exp}(-\frac{\chi}{k_B T_{\rm ion}}),
\end{equation}
where $\chi$
is the ionization energy.
When the emission originates
from a single-component, 
optically thin plasma in
CIE, 
the ionization temperature,
$T_{\rm ion}$, 
should match the excitation 
temperature, $T_{\rm exc}$. 
For Abell 2029, 
we calculate $T_{\rm ion}$ 
using the line flux ratios 
of Ly$\alpha$/He$\beta$, 
Ly$\alpha$/He$\gamma$, and 
Ly$\alpha$/z of Fe, 
as described in 
this study.

\section{Results}\label{sec:results}
Below we discuss results obtained from
spectral fitting.

\subsection{ICM properties}
\label{sec:icm_properties}
The ICM properties of the
central region of A2029 were 
previously reported in Paper I 
\citep{A2029_eric},
where the results were 
obtained without accounting
for spatial-spectral mixing. 
However, even after
incorporating SSM,
our measurements of the
ICM properties in the central
region remain consistent with 
those presented in Paper I. 
This consistency highlights 
that the spectrum from the 
central region is predominantly 
influenced by X-ray 
photons originating 
from that region, 
as clearly demonstrated 
in Figure \ref{fig:best_spec}.

\begin{figure}
    \includegraphics[width=0.47\textwidth]{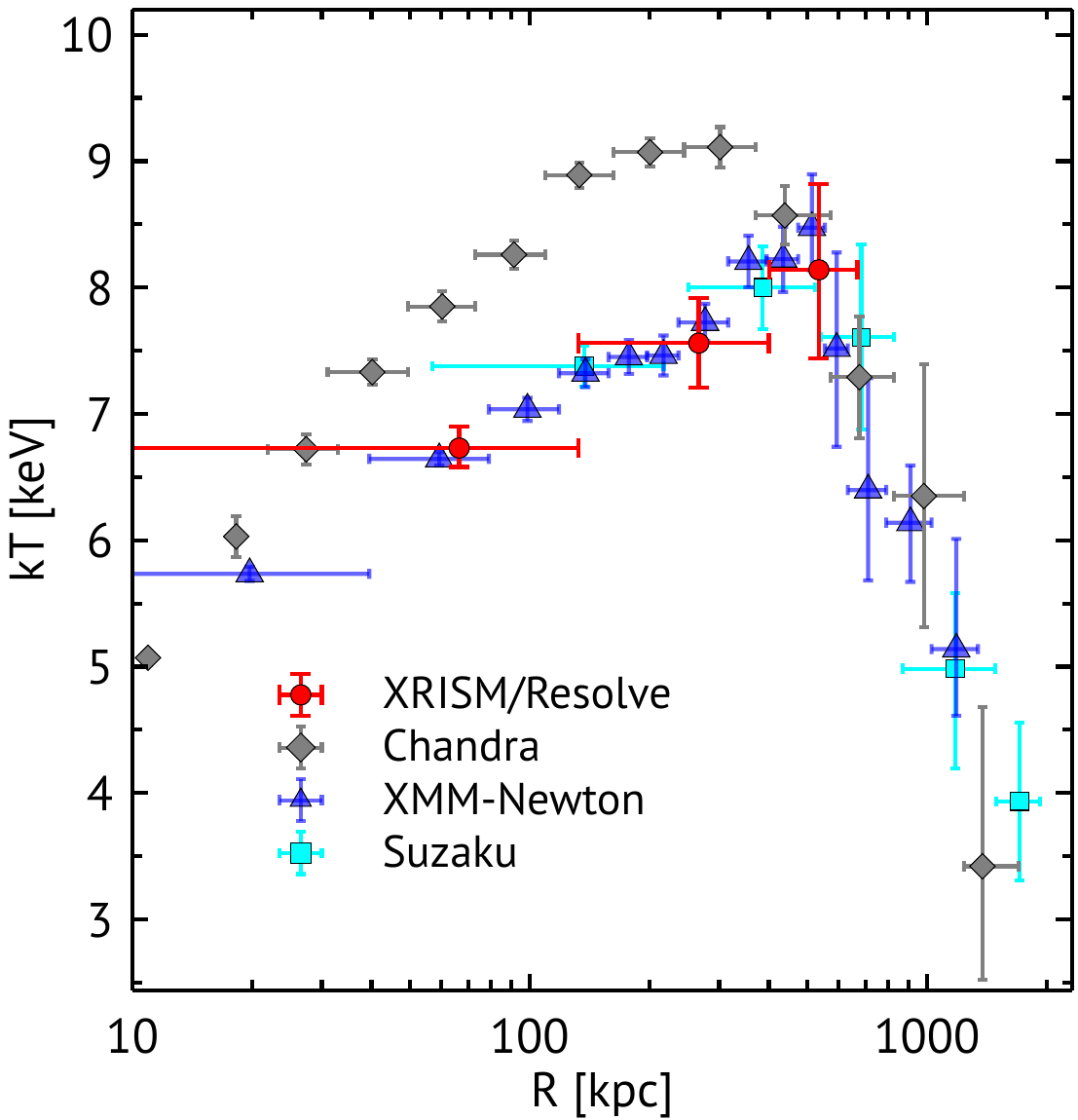}
    \caption{Comparison of the projected temperature profiles of A2029 as measured by XRISM/Resolve (red circles) with previous measurements from Chandra (gray diamond; \cite{2005ApJ...628..655V}), Suzaku (cyan square; \cite{Walkeretal2012a}),
    and XMM-Newton (blue triangle; 
    \cite{2019A&A...621A..41G}).
    {Alt text: Temperature profiles of A2029.}
    }
    \label{fig:temp_profile}
\end{figure}

\begin{figure}
    \includegraphics[width=0.47\textwidth]{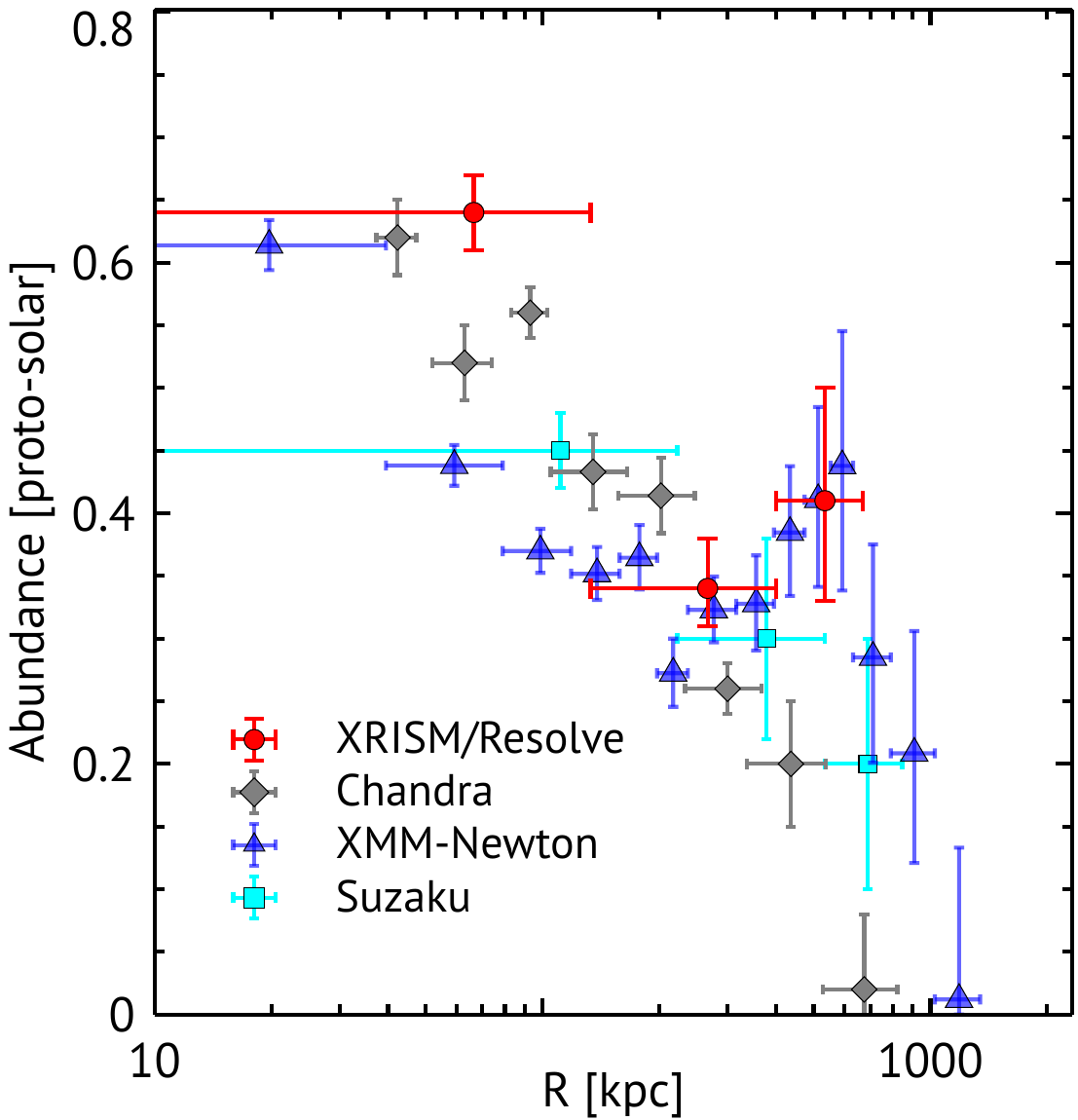}
    \caption{Elemental abundance profiles of A2029. The red circles represent measurements derived from Resolve data, while the gray diamonds, cyan squares, 
    and blue triangles correspond to abundance profiles obtained from Chandra (\cite{2005ApJ...628..655V}), Suzaku (\cite{Walkeretal2012a}), and
    XMM-Newton (\cite{2019A&A...621A..41G})
    observations, respectively. 
    {Alt text: Elemental abundance profiles of A2029.}}
    \label{fig:abun_profile}
\end{figure}

For the central region of A2029, 
we determine an average
gas temperature
of $6.73^{+0.17}_{-0.15}$ keV, 
which aligns with
previous measurements made 
using Suzaku \citep{Walkeretal2012a} 
and Chandra 
\citep{2005ApJ...628..655V}
within the central $<$2.5$\arcmin$. 
We also measure an
elemental abundance of 
$0.64^{+0.03}_{-0.03}$ $Z_{\odot}$ 
at the cluster center, 
consistent with earlier results. 
Figure \ref{fig:temp_profile}
and \ref{fig:abun_profile}
shows the resulting temperature
and abundance profiles.
Additionally, we find a best-fit 
redshift of 
$0.07776^{+0.00004}_{-0.00003}$ 
and a velocity broadening of 
$157^{+12}_{-13}$ km/s.

For the N1 and N2 regions,
we determine gas temperatures 
of $7.61^{+0.35}_{-0.33}$ keV 
and $8.14^{+0.70}_{-0.65}$ keV, 
respectively,
using single-temperature 
{\tt bapec} models. 
The N1 region spans a 
distance of approximately 
130–400 kpc from the center 
of A2029, 
while the N2 region 
lies at 400–670 kpc. 
Comparing our measurements
with previous observations 
from Suzaku and Chandra, 
we find that the temperature 
measurement for the N1 
region is 
{ significantly lower 
than previous Chandra 
measurements but aligns
well with that of
Suzaku and 
XMM-Newton}, 
as shown in Figure 
\ref{fig:temp_profile}. 
Similarly, for the N2 region, 
Resolve provides temperature 
measurements that align 
with those from 
Suzaku, Chandra, and 
XMM-Newton measurements. 
Figure \ref{fig:temp_profile} 
presents the temperature
profile of A2029, 
alongside previous 
measurements.

For the elemental abundances, 
we measure $0.34^{+0.04}_{-0.04} \, 
Z_{\odot}$ 
in the N1 region and 
$0.41^{+0.08}_{-0.07} \, 
Z_{\odot}$ in the N2 region. 
The abundance measurement for
N1 is consistent with both 
Suzaku, Chandra,
and XMM-Newton results. 
However, the N2 abundance is 
significantly
higher compared 
to Chandra measurements but 
remains consistent with
Suzaku and XMM-Newton
results. 
Figure \ref{fig:abun_profile} 
shows the abundance profile
of A2029 along with previous observations.
The best-fit redshifts for
N1 and N2 regions 
are $0.07706^{+0.00004}_{-0.00008}$ 
and $0.07788^{+0.00012}_{-0.00010}$, 
respectively. 
The corresponding line-of-sight
bulk velocities are reported
in Paper II \citep{A2029_naomi}.
The best-fit velocity broadening
($\sigma_{\rm v}$) for N2 is
is 92 km/s and the upper-limit for
N1 region is $<$82 km/s. 
A detailed study of gas velocities
is beyond the scope of this paper 
and are addressed in 
Paper II \citep{A2029_naomi}.
Table \ref{tab:best_params}
summarizes the best-fit
parameters for the central,
N1, and N2 regions.

{
Several previous studies 
have demonstrated the 
multiphase nature of ICM
(e.g., \cite{2022A&A...661A..36S,2023A&A...675A.150Z}). 
Accounting for this multiphase
gas is essential, 
as accurate ICM temperature 
measurements are crucial for 
determining cluster mass, 
abundance, and velocity dispersion.
Previous Chandra 
and XMM-Newton
observations have revealed a 
distinct temperature gradient 
within R$_{2500}$ of A2029 (e.g., 
\cite{2013ApJ...773..114P,2005ApJ...628..655V,2019A&A...621A..41G}),
which is covered by central, N1,
and N2 Resolve FOVs.
Given that the Resolve FOV is 
3 $\times$ 3 arcmin wide,
a single-temperature plasma 
model cannot adequately
represent the X-ray emission 
from these regions. 
}
To account for this, 
we employed a two-temperature 
(2T) CIE model as a
simplified approximation of 
deviations from the single-
temperature assumption. 
For the additional CIE component,
the temperature and normalization 
were allowed to vary, 
while the abundance, redshift, 
and $\sigma_{\rm v}$ were 
tied to those of the 
other temperature 
component. 

In the central region, 
the best-fit cooler component 
temperatures derived from 
AtomDB and SPEXACT are 
5.14$^{+0.52}_{-1.20}$ 
and 4.15$^{+1.14}_{-1.31}$, 
respectively. 
The hotter component temperatures 
are 9.19$^{+1.30}_{-2.32}$ and 
8.01$^{+1.87}_{-1.18}$,
respectively, as shown in
Table \ref{tab:best_params}. 
While SPEXACT yields gas 
temperatures that are 
approximately 20\% and 13\%
lower for the cooler and 
hotter components compared 
to AtomDB, these values 
remain consistent within 
their respective 1$\sigma$ 
uncertainties.

Similarly, for the N1 region
AtomDB yields a best-fit 
cooler component gas temperature 
of 3.58$^{+0.52}_{-1.20}$ and
a hotter component gas temperature 
of 8.61$^{+1.30}_{-2.32}$,
as shown in
Table \ref{tab:best_params}.
In contrast, with SPEXACT, 
it was not possible to constrain 
both temperatures. 
Instead, an average gas
temperature of 7.62 keV 
was obtained, 
consistent with a 
single-temperature plasma model. 
For the N2 region,
the X-ray spectrum is 
best represented by a 
single-temperature plasma
model using both atomic databases. 
In all three regions, 
the best-fit abundances, 
redshifts, and velocity
broadening obtained with both
AtomDB and SPEXACT are
consistent with the results
from single-temperature (1T) 
CIE models, as listed in 
Table \ref{tab:best_params}.

\subsection{Emission line-flux measurements}\label{sec:emission_flux}
The superb spectral resolution 
of XRISM Resolve enables,
for the first time, 
the detection of several key 
Fe and Ni emission lines 
in the central region and 
at two outer pointings of A2029. 
Figure \ref{fig:best_spec} 
presents the extracted spectra 
from the three Resolve FOVs, 
including zoomed-in views 
of the $\fexxv$, $\fexxvi$, 
and $\nixxvii$ emission lines
within the 6.5–8.4 keV (rest-frame) energy band.

In addition to the He$\alpha$ 
and Ly$\alpha$ emission lines 
of Fe, 
transitions from higher
principal quantum numbers are 
also resolved at the cluster center, 
such as $\fexxv$ He$\beta$,
$\fexxv$ He$\gamma$, 
and $\fexxvi$ Ly$\beta$. 
To measure the observed fluxes 
of these lines, 
we fitted the spectra in the
6.4–8.4 keV range using a 
continuum emission model 
(a no-line APEC model, {\tt nlapec}, 
combined with the NXB model) 
and a series of Gaussian components. 
The Gaussian line centers were
fixed to the rest-frame line 
centroids listed in Table 
\ref{tab:gaussian_lines},
scaled by the best-fit 
redshift provided in Table 
\ref{tab:best_params}. 
As with the Hitomi spectrum
of Perseus, $\fexxv$ He$\beta$1 
and $\fexxv$ He$\beta$2 are 
not resolved for A2029.
Consequently, their line 
centroids and widths were
tied during the fit. 
The unresolved line complex
of $\fexxv$ He$\gamma$ was 
modeled with a single
Gaussian component. 
The best-fit line fluxes 
and uncertainties were then
obtained using the
{\tt cflux} command in {\tt XSPEC}.

\begin{table*}
\caption{Line fluxes of bright lines using Gaussian line fitting\label{tab:gaussian_lines}}   
\begin{center}
\setlength{\tabcolsep}{6pt}
\begin{tabular}{lcccc}
Ions & Line E$_{0}^{\dagger}$ & Central & N1 & N2\\ 
& (keV) & ($10^{-15}$ erg s$^{-1}$ cm$^{-2}$) & ($10^{-15}$ erg s$^{-1}$ cm$^{-2}$) & ($10^{-15}$ erg s$^{-1}$ cm$^{-2}$)\\
\hline
\hline
$\fexxv\ z$ & 6.637 & 159.4 $\pm$ 18.8 & 16.8 $\pm$ 3.0 & 1.92 $\pm$ 0.23\\
$\fexxv\ x$ & 6.668 & 179.7 $\pm$ 20.7 & 19.5 $\pm$ 2.9 & 1.73 $\pm$ 0.21\\
$\fexxv\ y$ & 6.682 & 82.3 $\pm$ 14.8 & 5.28 $\pm$ 1.04 & 1.55 $\pm$ 0.24\\
$\fexxv\ w^{\dagger\dagger}$ & 6.700 & 503.6$^{+85.6}_{-25.2}$  & 55.9 $\pm$ 3.4 & 8.33 $\pm$ 0.42\\
$\fexxvi$ Ly$\alpha$1 & 6.952 & 128.9 $\pm$ 19.8 & 25.1 $\pm$ 4.8 & 3.67 $\pm$ 0.78\\
$\fexxvi$ Ly$\alpha$2 & 6.973 & 196.3 $\pm$ 21.9 & 31.6 $\pm$ 5.8 & 3.94 $\pm$ 0.56\\
$\fexxv$ He$\beta$ & 7.882 & 66.5 $\pm$ 7.9 & 8.28 $\pm$ 1.83 & 1.08 $\pm$ 0.13\\
$\fexxvi$ Ly$\beta$ & 8.253 & 57.8 $\pm$ 9.1 & $<$5.95 & 2.42 $\pm$ 0.88\\
$\fexxv$ He$\gamma$ & 8.295 & 26.3 $\pm$ 4.8 & $<$5.96 & $<$2.1\\
\hline
\end{tabular}
\end{center}
$^{\dagger}${\footnotesize Rest-frame wavelengths were 
taken from AtomDB database, using
their online waveguide.}
$^{\dagger\dagger}${\footnotesize A systematic uncertainty was added to the $\fexxv\ w$ line-flux to account for the effects of resonance scattering in the central region, as detailed in Appendix A.}
\end{table*}

The line fluxes obtained using 
this method also include 
contributions from other 
ionization states of Fe,
such as $\fexxiv$. 
To isolate the fluxes
corresponding to the desired 
ionization states,
we estimated the line 
fluxes using a full {\tt bapec} 
model, in which the intrinsic 
line emissivities of $\fexxv$
and $\fexxvi$ were set to zero.
{ All model parameters, 
including temperature, redshift, 
velocity broadening, 
and abundance, were fixed to 
the best-fit values listed 
in Table \ref{tab:best_params}.
The difference between the
line fluxes derived from the
initial method 
({\tt nlapec + 
Gaussians}) and the 
{\tt bapec} model 
(with $\fexxv$
and $\fexxvi$ emissivities zero) 
gives
the fluxes of individual 
ionization states of Fe.} 
These fluxes, along with 
their 1$\sigma$ uncertainties, 
are presented in 
Table \ref{tab:gaussian_lines}.

\subsection{Temperature via line ratio diagonistics}\label{sec:temp_via_line}
\begin{figure*}
    \begin{tabular}{ccc}
  \includegraphics[width=0.33\textwidth]{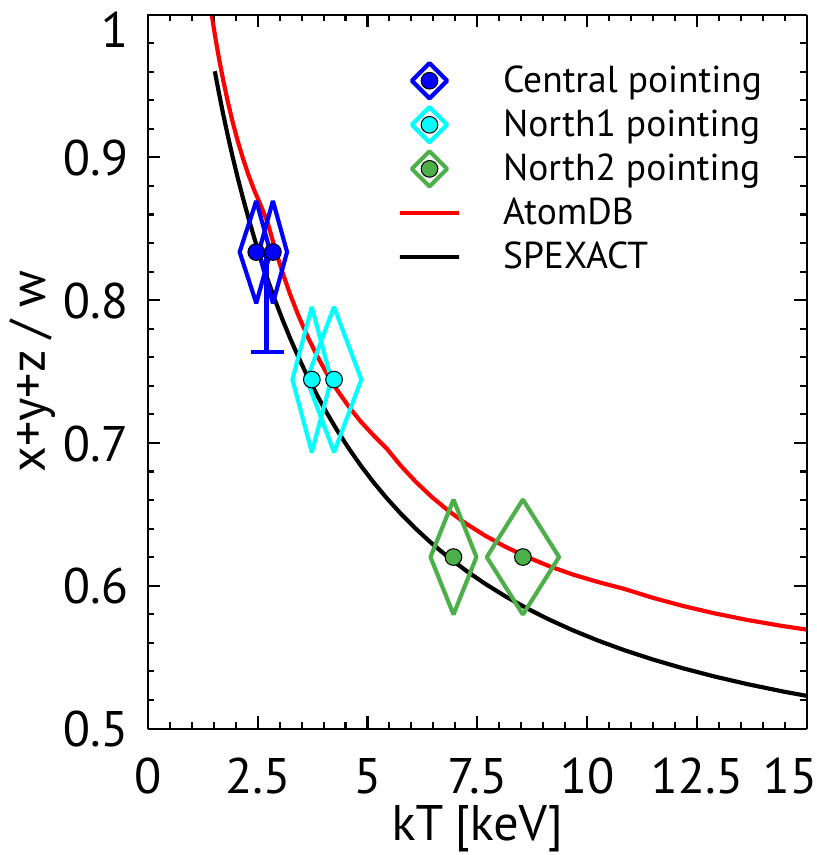} &  \includegraphics[width=0.33\textwidth]{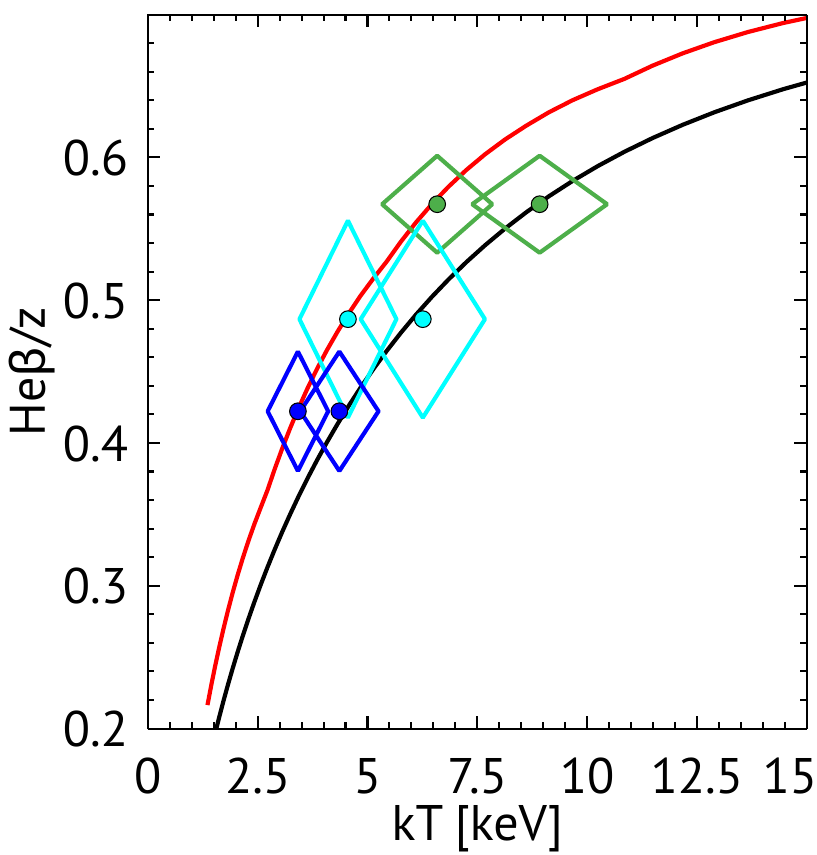}
         & 
 \includegraphics[width=0.33\textwidth]{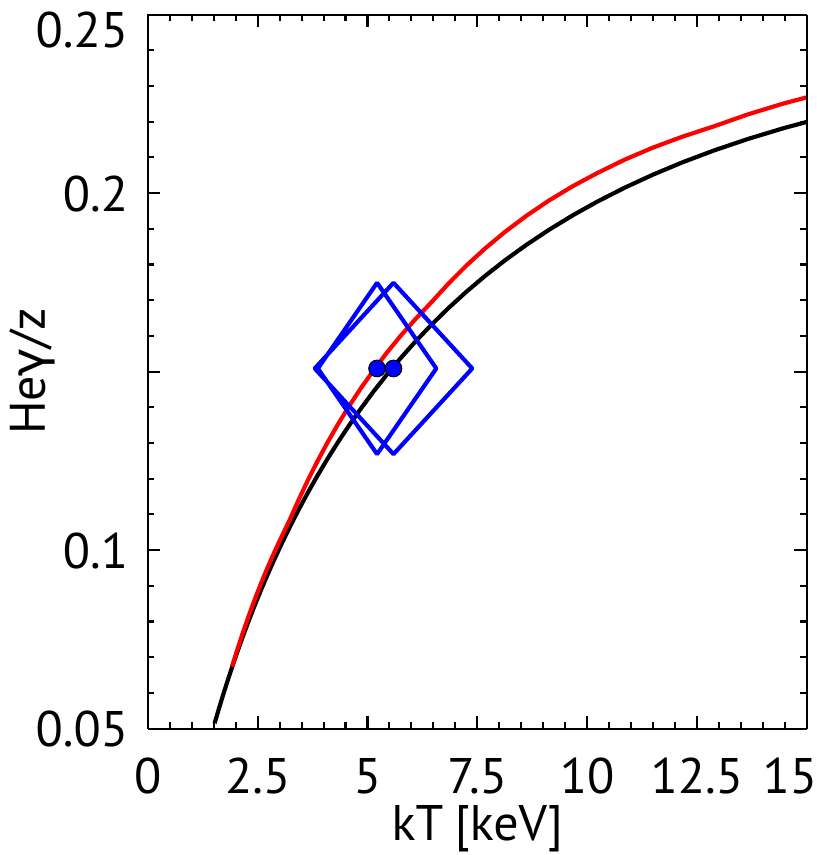}
    \end{tabular}
     \begin{tabular}{ccc}
  \includegraphics[width=0.33\textwidth]{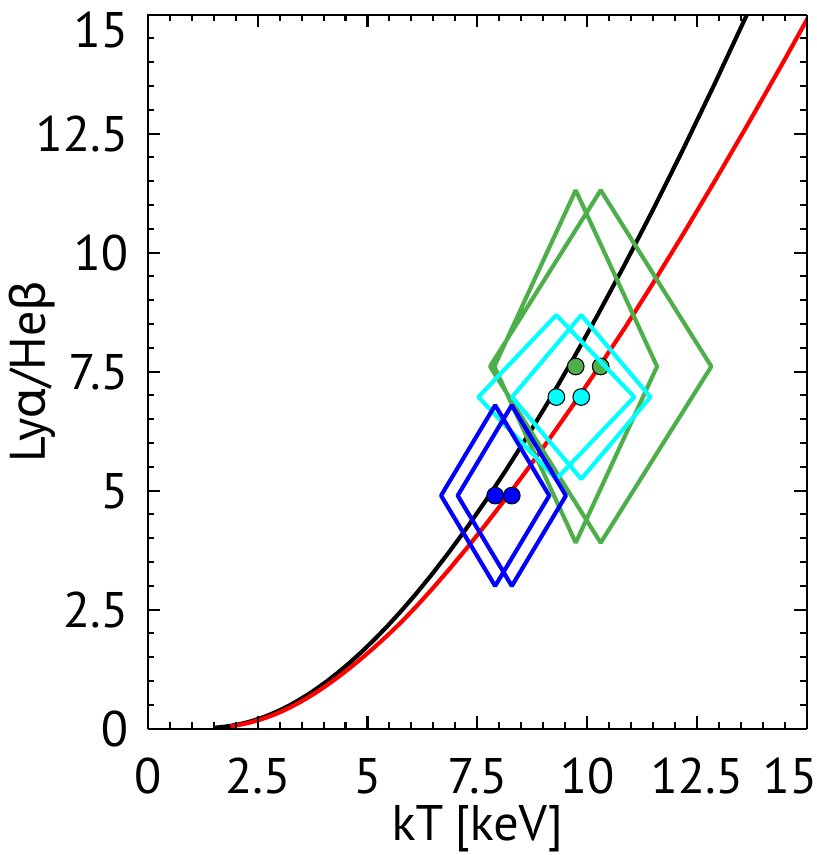} &  \includegraphics[width=0.33\textwidth]{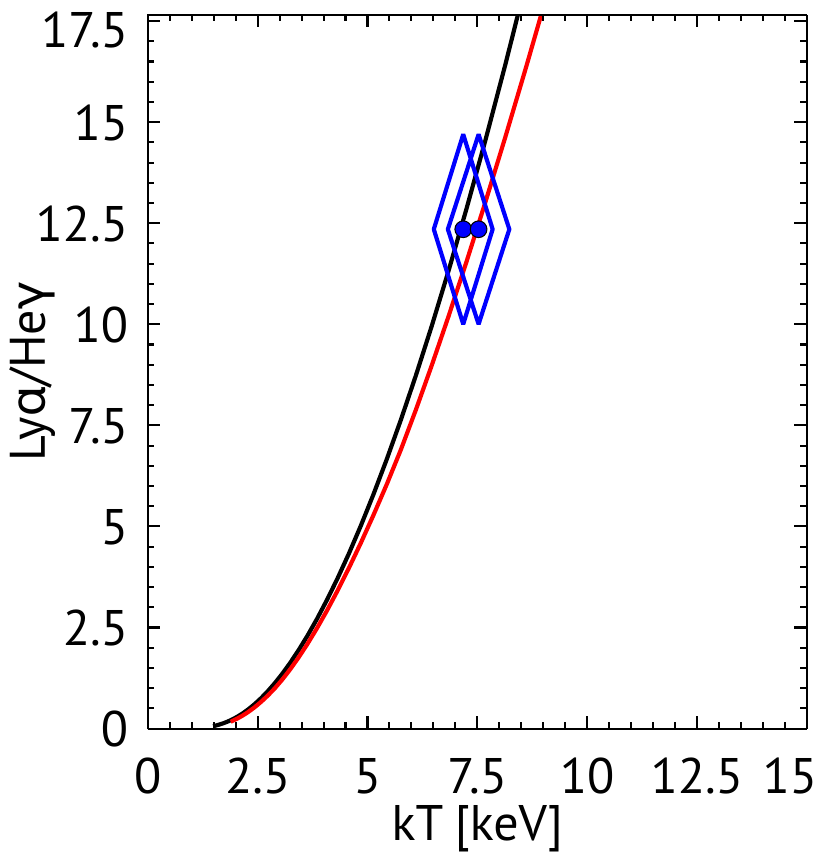}
         & 
 \includegraphics[width=0.33\textwidth]{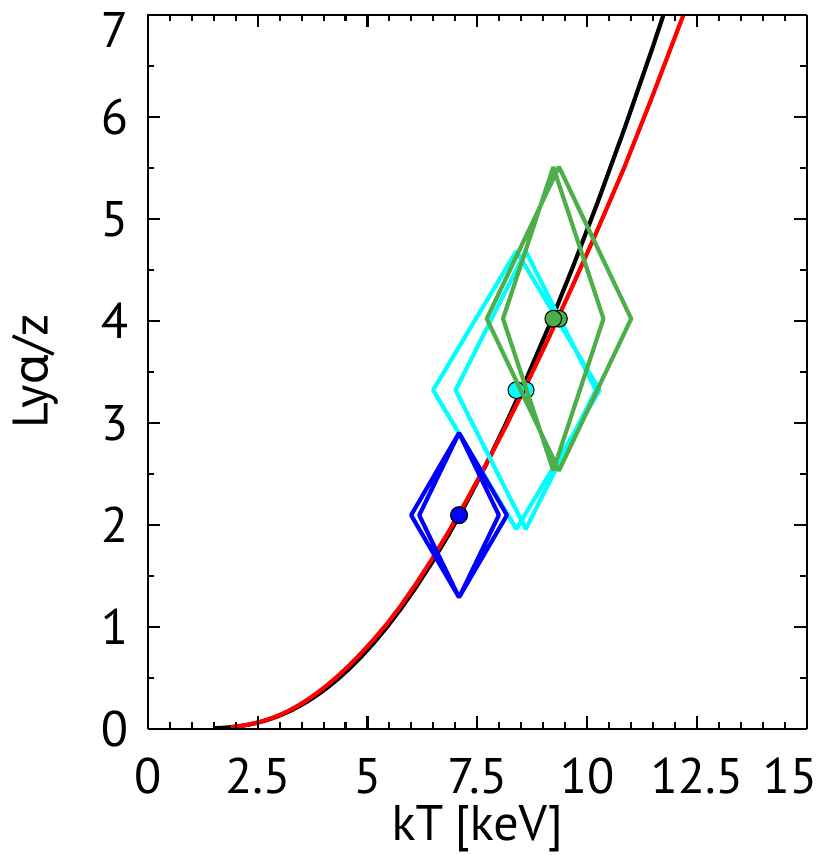}
    \end{tabular}
    \vspace{0.1in}
    \caption{Emission-line flux ratios of various Fe lines as a function of temperature, derived from the AtomDB (red) and SPEXACT (black) databases, assuming a single-temperature CIE plasma. The observed line ratios and their corresponding temperatures with 1$\sigma$
    uncertainties, as indicated by the AtomDB and SPEXACT curves, are presented with blue (central pointing), cyan (North1 pointing), and green (North2 pointing) circles, respectively.
    { The downward blue error bar in the 
    $x+y + z / w$ ratio indicates the 
    1$\sigma$ limit of the systematic 
    uncertainty due to resonance scattering.}
    {Alt text: Predictions of different key line-ratios from two atomic databases.}
    }
    \label{fig:line_ratios}
\end{figure*}

In the spectral analysis described 
in Section 
\ref{sec:spectral_analysis}, 
the shape of the bremsstrahlung 
continuum predominantly drives
the temperature measurement
across broad energy bands. 
{
Continuum-derived
temperatures may be different
from the actual ICM temperatures
if the X-ray emission is 
impacted by non-equilibrium
ionization arise from any prior
merging events 
\citep{2010A&A...509A..29P}.
}
Also, the ability to 
identify the presence of
multi-temperature gas in 
galaxy clusters is often 
limited by the statistical 
quality of the observed data,
as the spectral fit also 
depends on parameters 
such as elemental abundance. 
To address these limitations, 
we adopt an alternative and 
well-established method: 
line-ratio diagnostics, 
to probe multi-temperature gas
in A2029 
\citep{2010A&A...523A..22N,2018PASJ...70...11H}. 
In this approach, 
the observed flux ratios of 
temperature-sensitive
emission lines are compared 
to the ratios predicted by the 
{\tt AtomDB} and {\tt SPEXACT} 
databases. 
Figure \ref{fig:line_ratios} 
illustrates the predicted 
line ratios from these databases 
as a function of gas temperature. 
The observed line ratios 
for the different
Resolve pointings were 
derived using the line fluxes
listed in Table 
\ref{tab:gaussian_lines} 
and discussed in the
previous section.  
The two databases predict 
consistent flux ratios within 
a 5–10\% range.

\begin{figure*}
\begin{center}
    \begin{tabular}{cc}
    \includegraphics[width=0.5\textwidth]{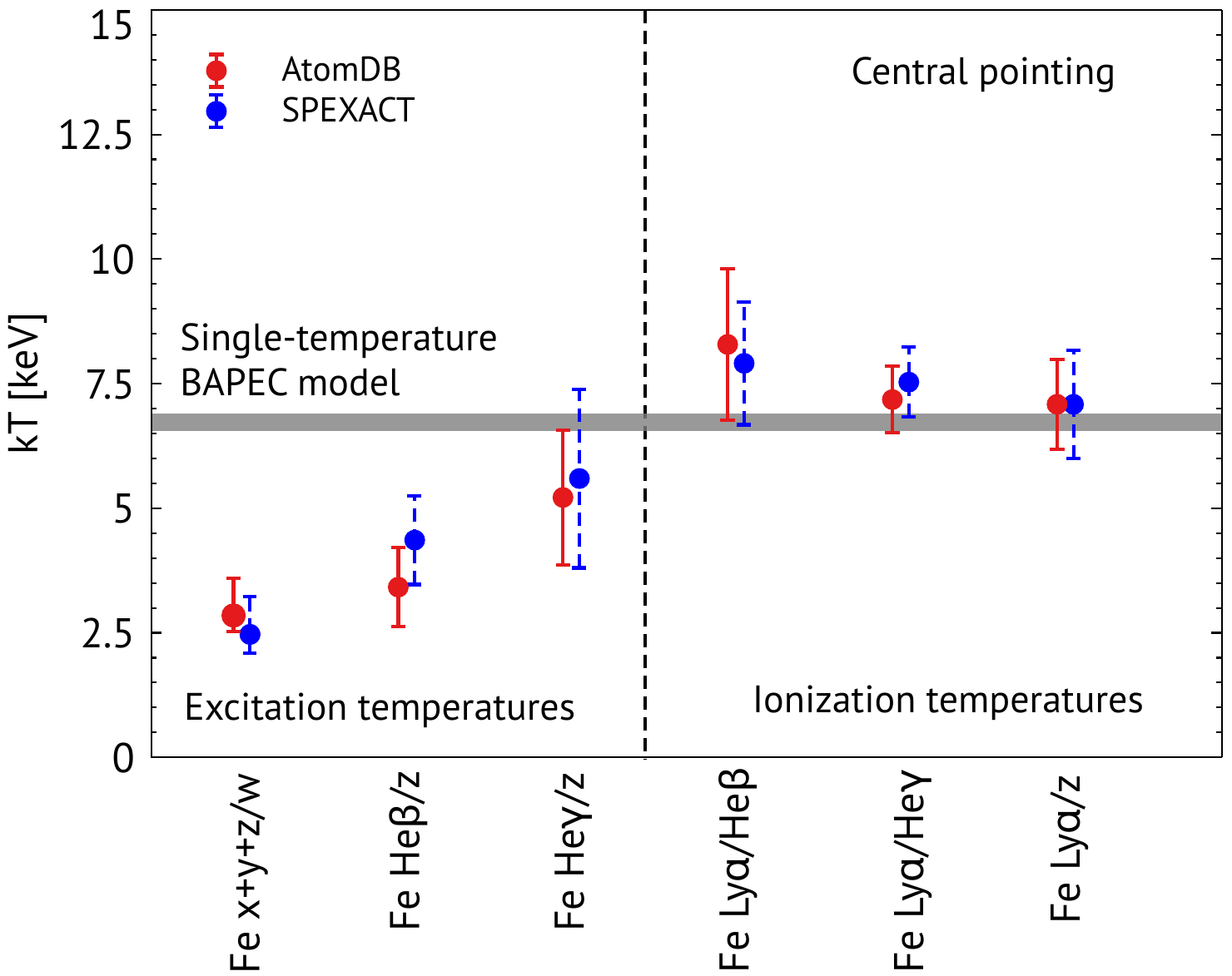} & \includegraphics[width=0.5\textwidth]{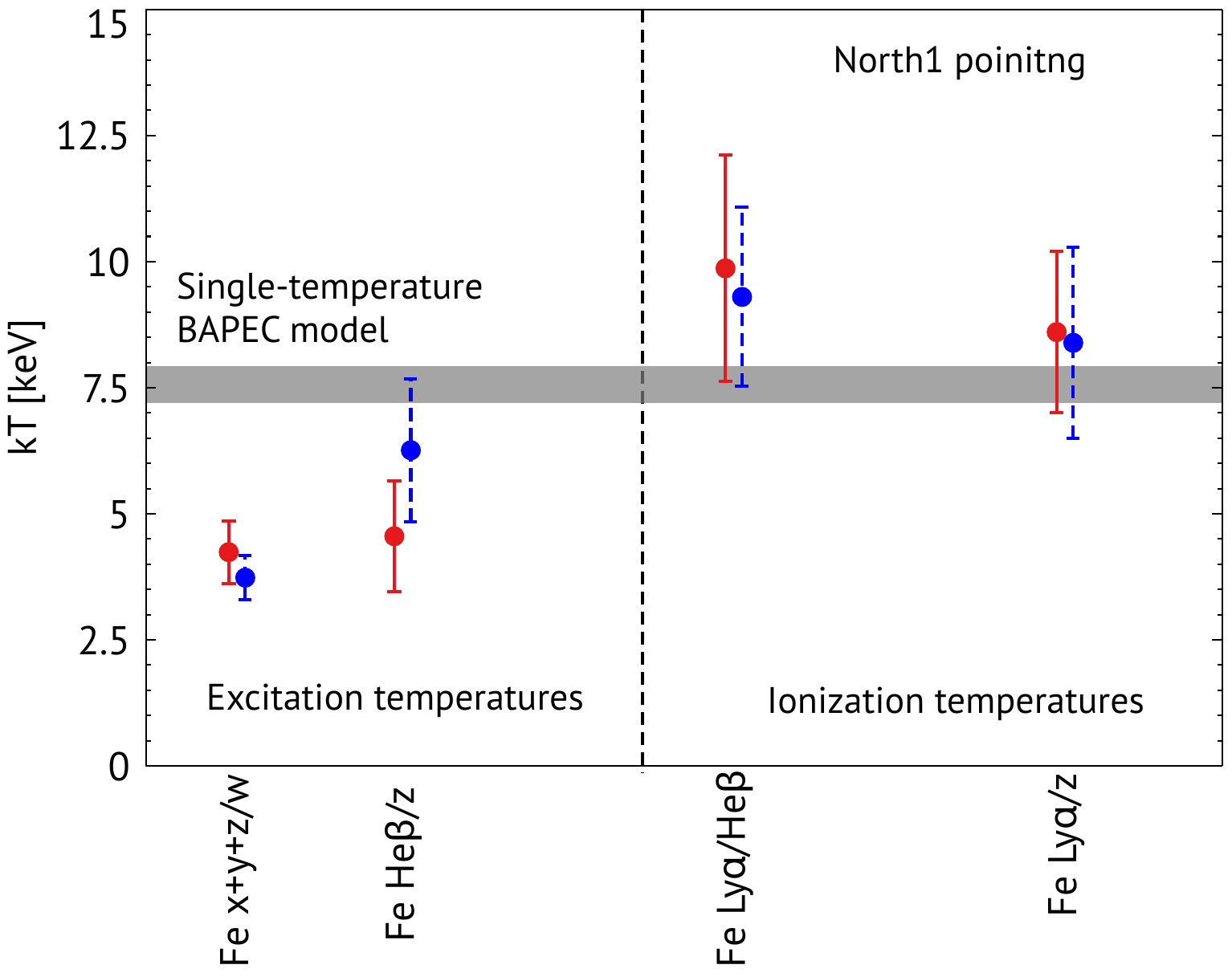}
    \end{tabular}
    \begin{tabular}{c}
\includegraphics[width=0.5\textwidth]{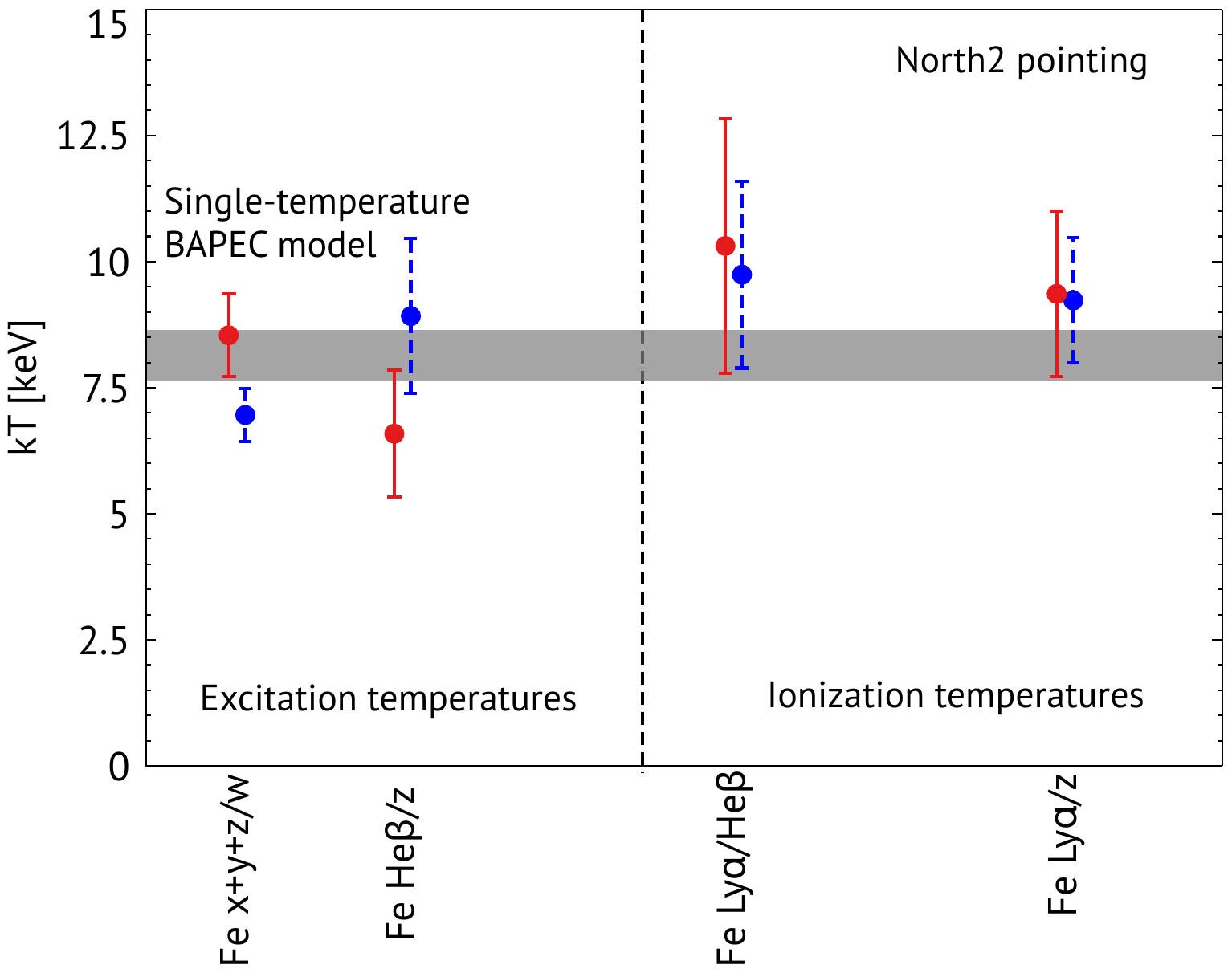}
    \end{tabular}
\end{center}
    \vspace{0.1in}
    \caption{Excitation and ionization temperatures of A2029 derived from emission-line flux ratios of Fe. Red and blue points represent temperatures calculated using the AtomDB and SPEXACT databases, respectively, with the blue data points shifted slightly to the right for better visibility. 
    { A systematic uncertainty (1$\sigma$ upper-limit) is added to the
    temperature inferred from the Fe $x+y+z/w$ 
    line in the central pointing, in addition to the statistical uncertainties.}
    In all three panels, the gray-shaded horizontal lines indicate the best-fit temperatures from the single-temperature models, as listed in Table \ref{tab:best_params}.
    {Alt text: Excitation and ionization temperatures inferred from line-ratio diagnostics.}}
    \label{fig:temp_lineratio}
\end{figure*}

We focus on measuring the 
excitation ($T_{\rm exc}$) 
and ionization ($T_{\rm ion}$) 
temperatures for the central, 
N1, and N2 pointings of A2029. 
For a detailed discussion on 
the underlying physics of 
excitation and ionization 
temperatures, we refer readers 
to Section \ref{sec:physics_line}. 
The excitation temperature,
$T_{\rm exc}$, is derived by 
comparing the observed line
flux ratios—He$\beta$/z, 
He$\gamma$/z, and $x+y+z/w$ of
Fe—with predictions 
from atomic databases.
Since He$\gamma$ is detected
with a significance of
$\geq$ 3$\sigma$ only in the
central region, 
it is not utilized for the N1 
and N2 pointings.
Additionally, the fluxes 
of the fine structures of
He$\beta$ and He$\gamma$ are summed,
as these lines are unresolved. 
The ionization temperature, 
$T_{\rm ion}$, 
is determined using the line 
ratios Ly$\alpha$/He$\beta$, 
Ly$\alpha$/He$\gamma$, 
and Ly$\alpha$/z for Fe. 
For the N1 and N2 pointings,
the Ly$\alpha$/He$\gamma$ ratio
is excluded due to its 
non-detection. 
Figure \ref{fig:temp_lineratio} 
presents the resulting
excitation and ionization 
temperatures for the central, 
N1, and N2 Resolve FOVs.

If the X-ray emission of A2029 
were dominated by a 
single-component, 
optically thin plasma under 
collisional ionization equilibrium, 
the ionization and excitation 
temperatures would be consistent
with each other. 
However, our results indicate that, 
in all three regions,
the excitation temperatures
differ significantly from 
the ionization temperatures,
except for Fe He$\gamma$/z in
the central region and
Fe $x+y+z/w$ in the N2 region.
In the central region,
the excitation temperatures 
range from 2.8 to 6.4 keV,
which are significantly lower
than those predicted by the
single-temperature plasma model, 
with deviations at a confidence
level of $\geq$ 5$\sigma$ 
(excluding Fe He$\gamma$/z). 
This discrepancy highlights
a clear deviation from a 
single-temperature plasma model. 
In contrast, the ionization 
temperatures are mostly 
consistent with the continuum 
temperature within their 
1$\sigma$ uncertainties.
Similarly, for the 
N1 and N2 regions, 
the excitation temperatures are 
lower than those predicted by 
single-temperature models,
with deviations at $\sim$ 
5$\sigma$ and 3$\sigma$ 
confidence levels, respectively 
(excluding Fe x+y+z/w for N2). 
In each case, the ionization 
temperatures remain consistent
with the single-temperature
models.

{ 
We note that our measured excitation
temperatures are lower than the ionization
temperature because they are derived
from $\fexxv$ line ratios,
whereas ionization temperatures are
determined using $\fexxv$/$\fexxvi$ 
line ratios. 
As shown in Figure 
\ref{fig:ionization_vs_temp},
at lower temperatures
($<$ 5 keV), 
the ionization fractions
of 
Fe XXV lines are more than 
twice 
as high as those of $\fexxvi$.
Consequently, temperatures measured 
from different excitation states of 
$\fexxv$ are more sensitive to 
lower-temperature gas. 
}

\subsection{Systematic Uncertainties}

We tested the results reported 
in Table \ref{tab:best_params} 
against potential systematic
biases arising from several factors,
including, 
CXB, NXB models, and 
gain calibrations (see Section 
\ref{sec:rslv_data_reduction}). 
Due to the design of
atomic databases,
electron temperature and 
abundances are typically 
measured by fitting the
broad-band continuum emission. 
These measurements are susceptible 
to systematic biases if 
the CXB and NXB are not
modeled accurately. 
In contrast, velocity 
broadening and redshift are
primarily derived from 
spectral lines and are
therefore less sensitive to 
variations in background levels.

First, we allowed the 
best-fit normalization of
the CXB power-law component 
to vary by $\pm$30\%,
while keeping the power-law 
slope fixed at $\Gamma = 1.41$ 
(\cite{2004A&A...419..837D}). 
For the central and N1 regions, 
variations in the CXB normalization 
had no significant impact on 
the best-fit parameters.
However, for the N2 region,
the abundance changed by
$\leq$ 7\% relative to the
best-fit values reported 
in Table \ref{tab:best_params}. 
For the 2T models, variations
in the CXB normalization had no 
significant impact on the best-fit 
parameters for the central and N2 
regions but altered the cooler gas 
temperature in the N1 region by 
approximately 14\%.
We also tested the impact of 
fixing the CXB power-law 
slope to $\Gamma = 1.3$ and 
$\Gamma = 1.5$ 
(\cite{2004A&A...419..837D,2021MNRAS.501.3767S}). 
These changes did not affect
the best-fit parameters for 
the single-temperature models 
in the central, N1, and N2 regions. 
However, for the 2T models, 
varying $\Gamma$ altered the
cooler temperature component 
in the N1 region by 25\%.
Since the two unabsorbed 
components in the sky-background 
model, 
as discussed in Section 
\ref{sec:sky_bkg}, 
do not contribute significantly in the
2--10 keV energy band, 
we excluded them from 
this analysis.

Another significant source 
of systematic uncertainty is 
the Resolve NXB, 
which has not yet been well characterized due to
XRISM’s early 
operational phase. 
However, the XRISM background
team has developed a detailed
model of the Resolve NXB, 
based on limited observations
of the night Earth 
(\cite{10.1117/12.3018882}), 
as discussed in Section 
\ref{sec:nxb_model}.
To evaluate the impact of 
these uncertainties, 
we varied the constant parameter
in the NXB model by $\pm$20\%. 
For the central and N1 regions,
these variations had no 
significant effect on the
best-fit parameters. 
However, in the N2 region, 
the temperature changed by 
approximately 7\%.
For the 2T models, 
the cooler and hotter
temperature components in
the N2 region changed by
$\sim$27\% and $\sim$10\%, 
respectively,
while the abundance 
varied by $\sim$6\%.

{ Another potential source of systematic uncertainty is resonant scattering in the optically thick He-like Fe ($\fexxv$ $w$) line, particularly in the central region. To minimize its impact on the best-fit parameters in the 1T and 2T spectral models, we replaced the $w$ line with a Gaussian profile, following \cite{2018PASJ...70...11H}. This substitution ensures that the results presented in Table \ref{tab:best_params} are not biased by resonant scattering. We further tested this by re-fitting the Resolve broad-band spectra after excluding the $w$ line using the {\tt XSPEC} ``ignore'' command. This exclusion did not affect the best-fit parameters, indicating that they are not significantly influenced by the presence of resonant scattering features.
{ To quantify the impact of resonant scattering on the estimated line flux—specifically the $\fexxv\ w$ line—we refitted the Resolve spectra using a single-temperature, absorbed {\tt RSAPEC} model \citep{2023ApJ...959..126C}, as described in Appendix A. Although the limited Resolve exposure prevents a direct detection of the resonant scattering signature in the $\fexxv\ w$ line, we derived a 1$\sigma$ upper limit on its suppression at the 12\% level. This potential suppression translates to a systematic uncertainty of $+$0.43 keV in the temperature inferred from the $\fexxv\ (x + y + z)/w$ line ratio, in addition to the statistical uncertainty of 0.32 keV. These systematic uncertainties on both the line-flux ratios and the inferred temperatures are illustrated in Figures \ref{fig:line_ratios} and \ref{fig:temp_lineratio}.}

{ Projection effects can also contribute to systematic uncertainties in the estimated line fluxes, particularly in the central region. To evaluate this, we adopted the approach described in \cite{2018PASJ...70...11H}. Specifically, we used azimuthally averaged, deprojected radial profiles of temperature, metal abundance, and electron density derived from XMM-Newton observations, covering radii from the cluster center out to 1300 kpc. These deprojected profiles were then integrated along the line of sight to generate a simulated Resolve spectrum for the central region, from which we estimated the expected line ratios. Figure \ref{fig:projection} shows a comparison between the observed and model-predicted line ratios for each line species. The close agreement between them suggests that projection effects do not significantly bias the measured line ratios in the central region.
}

\begin{figure}
    \includegraphics[width=0.45\textwidth]{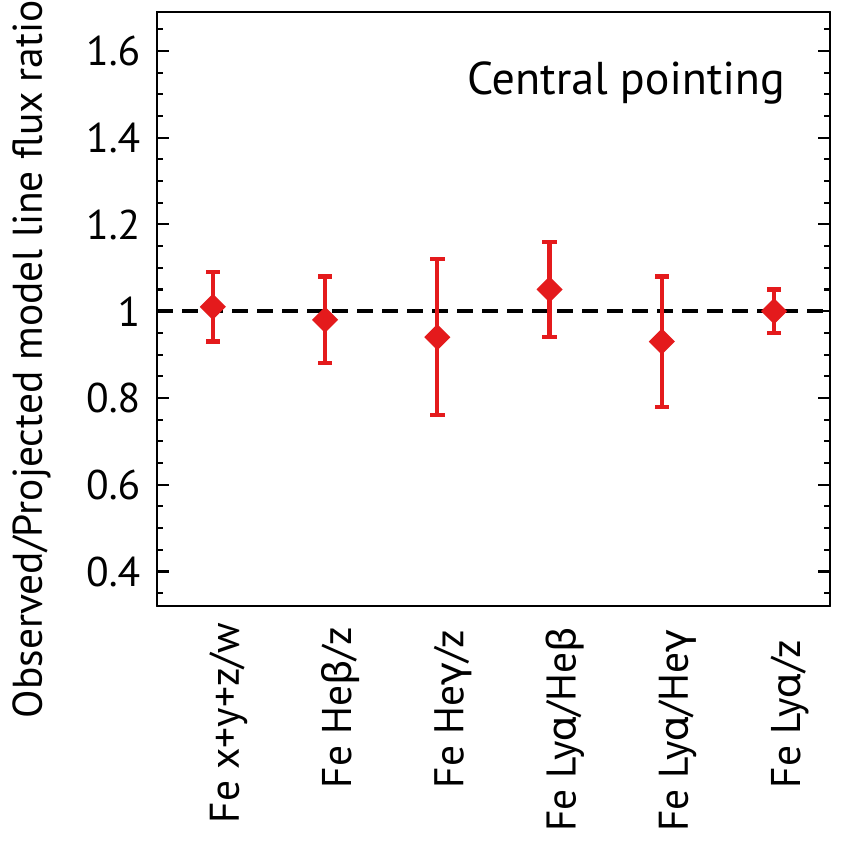}
    \caption{The ratio of the observed to the projection model in the central region. The vertical dashed line separates the line ratios used to estimate the excitation temperature (left side) from those used to estimate the ionization temperature (right side). {Alt text: The ratio of observed to projection model in the central region.}}
    \label{fig:projection}
\end{figure}

\section{Multi-temperature gas and its origin}\label{sec:origin}
ICM exhibits multi-phase gas
with
range of 
temperatures and metallicities,
which needs to be accounted for accurate
metal abundance and velocity dispersion
measurements.
However, gas temperature measured 
from continuum fitting 
fails to capture the multiphase 
structure of the gas, 
leading to biases in the 
estimated temperature and 
elemental abundance (e.g., 
\cite{2023A&A...675A.150Z,2004A&A...413..415K}). 
In low-exposure 
observations of galaxy clusters, 
there are typically insufficient 
counts to model the emission 
within a given projected radius
using anything beyond a 
single-temperature and
single-metallicity plasma 
emission model.
{ The Resolve microcalorimeter
provides a novel approach to 
studying the multi-phase ICM 
in galaxy clusters through
high-resolution spectral analysis.
Using Resolve observations, 
we identify multi-temperature
gas in the central and N1 
regions of A2029.
Below, we discuss these findings in detail, focusing on the nature of the observed multi-temperature structure and investigating the physical processes that may give rise to it.}

\subsection{Multi-temperature structures}
{ The Resolve FOV covers a radial
range of 0–1.5 arcmin in the 
central region of A2029. 
Within this radius, 
Chandra 
observations revealed 
temperatures ranging from 5.05 to 
8.89 keV (specifically 5.05, 
6.03, 6.72, 7.34, 7.85, 
8.25, and 8.89 keV) across seven 
radial bins 
(\cite{2005ApJ...628..655V}). 
Here, we perform a similar
multi-temperature analysis 
to assess the consistency
between XRISM observations
with previous observations.}
We fitted the Resolve spectra
from the central region assuming 
a seven-temperature (7T) 
plasma emission model. 
The temperature components 
were fixed to the corresponding 
values obtained from Chandra 
observations for each 
radial bin.
The velocity broadening and 
metal abundance were allowed 
to vary but were tied across
the individual components.
We utilized both the AtomDB 
and SPEXACT databases and 
compared the best-fit 
normalizations between the two.

Figure \ref{fig:chandra_temp_model} 
(left) shows the best-fit
models for the central region, 
with the individual temperature 
components marked in 
different colors. 
The best-fit normalizations 
for each component are scaled 
so that their sum equals unity. 
Figure \ref{fig:chandra_temp_model} 
(right) compares the resulting
scaled normalizations obtained
from AtomDB and SPEXACT with 
that of derived from Chandra observations
\citep{2005ApJ...628..655V}.
The scaled normalizations 
between AtomDB and SPEXACT are 
generally consistent, 
except for the 5.05 keV component, 
which is constrained using 
AtomDB but only has an upper 
limit when using SPEXACT.
These results indicate that
the two-temperature model discussed
in Section 
\ref{sec:icm_properties} 
provides an approximation of
the combination of 5.05 keV,
6.03 keV, 8.25 keV, and 
8.89 keV components.
{ Although the distribution of higher-temperature components in the XRISM observations
generally follows the trends
seen in the Chandra measurements
(Figure \ref{fig:chandra_temp_model} left), 
there is a noticeable discrepancy in the lower-temperature components. 
This difference may be attributed to the choice of fitting energy band in the Chandra/ACIS analysis
(0.6--10 keV), 
which influences the derived normalization ratio for
lower temperatures.
A similar conclusion was previously reported in the analysis of Hitomi observations of the Perseus cluster
\citep{2018PASJ...70...11H}.
We used the 7T
model mainly to check 
consistency with previous
measurements.
However, we recognize that
this approach may not fully 
capture the complexity of the ICM.
}

\begin{figure*}[h!]
\begin{center}
    \begin{tabular}{cc}
\includegraphics[width=0.5\textwidth]{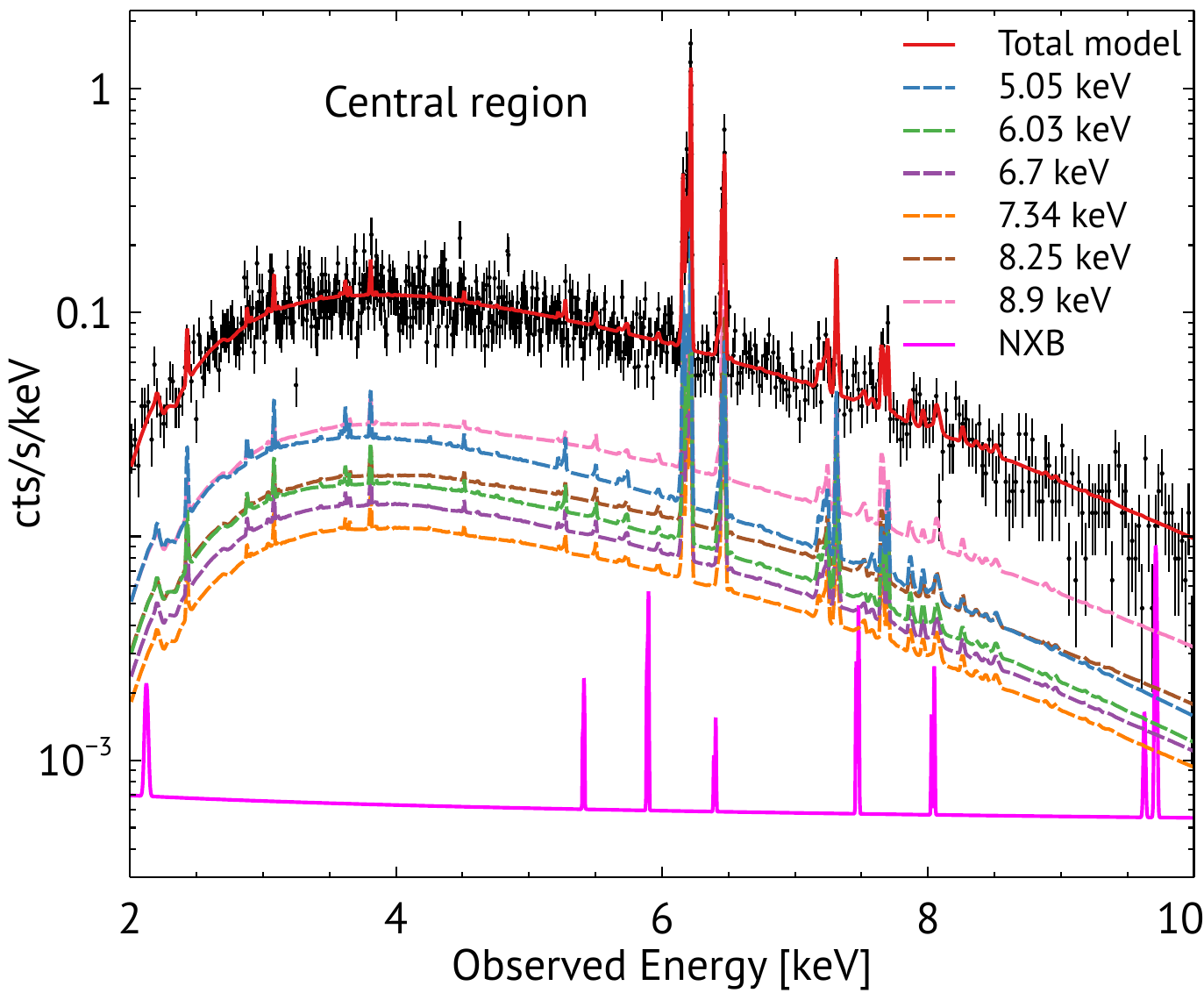} & \includegraphics[width=0.39\textwidth]{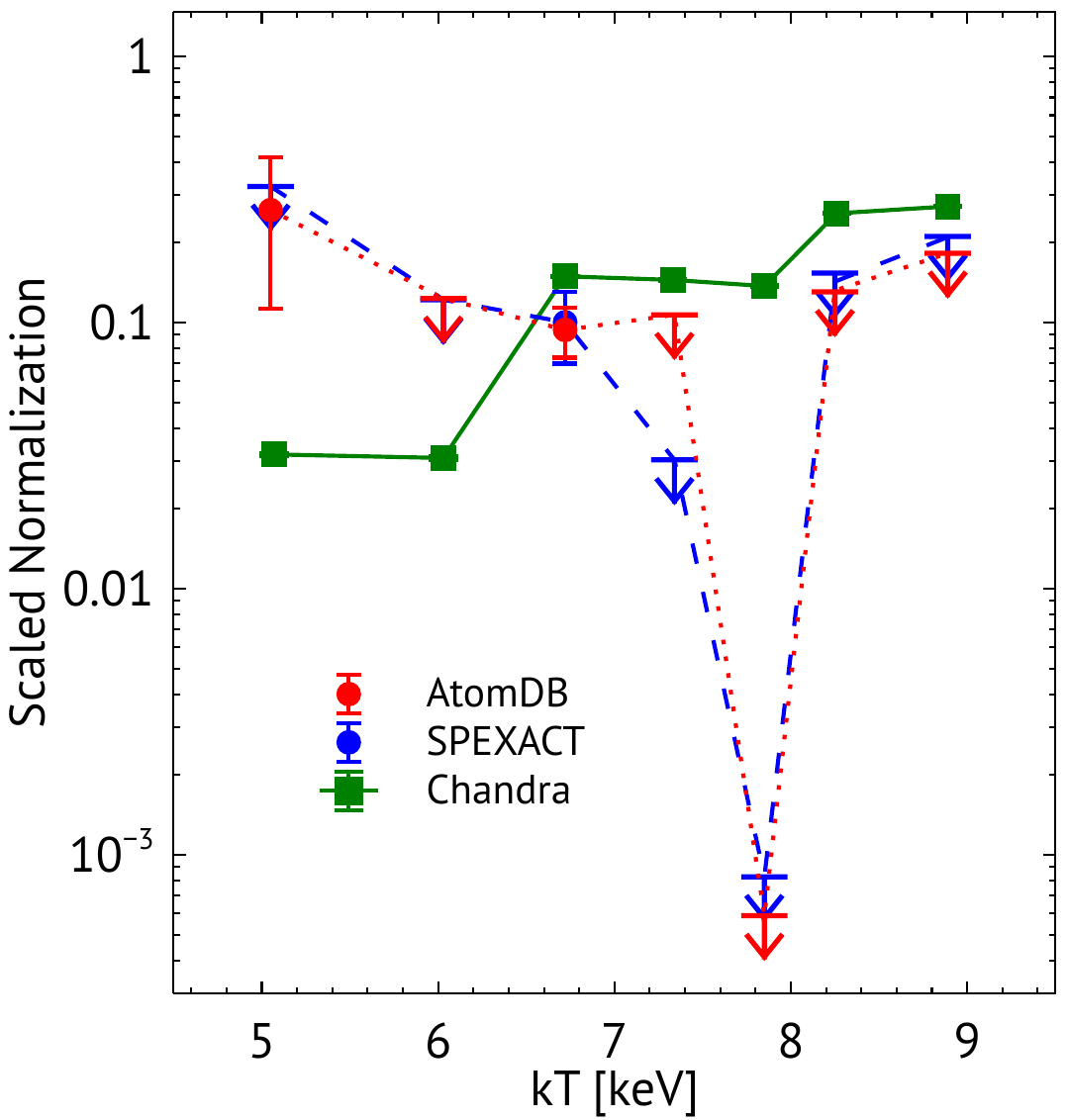}
    \end{tabular}
\end{center}
    \caption{Left: XRISM/Resolve spectrum of the central region of A2029 (black data points) fitted with a 7-temperature plasma model, where the temperatures are fixed to the Chandra temperature profile. The red curve represents the best-fit total model, while other colors correspond to individual temperature components. The magenta curve shows the NXB spectrum.
Right: Comparing scaled best-fit normalizations of each temperature component in the 7-temperature CIE model with the Chandra observations. 
Green squares indicate scaled
normalization derived from Chandra spectral fit \citep{2005ApJ...628..655V}.
Red and blue data points represent scaled normalizations derived using the AtomDB and SPEXACT databases, respectively.
{Alt text: Resolve spectrum from the central region is fitted with a seven-temperature plasma emission model.}
}
\label{fig:chandra_temp_model}
\end{figure*}

Our line-ratio diagnostics  
measurements using Resolve
also
reveal multi-temperature gas
in the central region,
as shown in Figure 
\ref{fig:temp_lineratio}. 
Cooler gas is detected at 
temperatures as low as 
2.85$^{+0.75}_{-0.32}$ keV from the
Fe $x+y+z/w$ line ratio,
while hotter gas is identified 
at 8.3 $\pm$ 1.5 keV from the
Fe Ly$\alpha$/He$\beta$ line ratio. 
Resolve also detects 
intermediate temperatures, 
including 3.42 $\pm$ 0.78 keV 
(Fe He$\beta$/z), 
7.09 $\pm$ 0.61 keV 
(Fe Ly$\alpha$/He$\gamma$), 
5.21 $\pm$ 1.35 keV 
(Fe He$\gamma$/z), 
and 7.08 $\pm$ 0.91 keV 
(Fe Ly$\alpha$/z). 
The temperatures of 5.21 keV
and 8.3 keV are consistent 
with the components of the 
2T model discussed in
Section \ref{sec:icm_properties}.
{ All Resolve temperature
measurements in the central
region align
with previous 
Chandra and XMM-Newton
observations, except for
the cooler 2.85 keV and 3.42 keV 
components.}
Notably, these two components 
are $>$11$\sigma$ and 5$\sigma$ 
cooler than the single-temperature 
plasma model, providing strong 
evidence for the presence of
multi-temperature structures 
in the central region of A2029. 
Both temperatures are 
marginally consistent
with the cooler component of 
the 2T plasma model 
discussed earlier.
{ We also attempted to fit
the broad-band Resolve spectrum from the
central region using 
a 3T plasma emission model
(as detailed in Appendix B), 
aiming to explore the possible contribution
of cooler-gas components to the 
overall emissivity.}

{ Furthermore, 
in the N1 region, Resolve
reveals multi-temperature gas,
with higher temperatures
derived from Fe 
Ly$\alpha$/z (8.6 $\pm$ 1.6 keV) 
and Fe Ly$\alpha$/He$\beta$ 
(9.8 $\pm$ 2.2 keV) line ratios,
as shown in Figure 
\ref{fig:temp_lineratio}.
These values are 
consistent with previous
Chandra and XMM-Newton 
observations.
Additionally, Resolve detects 
cooler gas in the
N1 region, with temperatures 
derived from 
Fe $x+y+z/w$ (4.24 $\pm$ 0.62 keV) 
and Fe He$\beta$/z 
(4.55 $\pm$ 1.10 keV) line ratios, 
as shown in Figure 
\ref{fig:temp_lineratio}. 
These temperatures are
significantly lower than the
single-temperature model
described in Section 
\ref{sec:icm_properties}, 
with deviations at a 
4.5$\sigma$ and 2.7$\sigma$ levels, 
respectively. 
Notably, 
these low-temperature components
are consistent with the cooler 
component of the two-temperature 
model reported in Table 
\ref{tab:best_params}.

\subsection{The Cooling Core}
The origin of   
multi-temperature
gas in the central and N1 regions, 
as revealed by Resolve, 
is linked to 
the complex thermal history
of A2029.
\cite{2004ApJ...616..178C} 
proposed that A2029 might either 
be a very young cooling 
core cluster, 
where the gas is cooling
for the first time, 
or a system where a previous 
cooling episode was disrupted
by a merger event and/or 
activity from the central AGN.
Both scenarios are unlikely as Abell 2029 is a relatively relaxed cluster with an enormous, well-established cooling core extending to $\sim 100~\rm kpc$.  
The central region of A2029
features a 
central Wide-Angle Tail (WAT)
AGN and X-ray gas 
sloshing spiral 
(Figure \ref{fig:chandra_image}), 
both of which  
indicate AGN feedback amidst gas flows from a past 
minor merger
(\cite{2004ApJ...616..178C}). 
These processes will stir and heat the atmosphere, but they are not associated with a destructive event.

The temperature components
resolved: 5.21, 7.08, 7.09, 8.3 keV
in the central
region, 8.6 and 9.8 keV 
in the N1 region,
and $\sim$ 8 keV 
in the N2 region--
are consistent with 
previously observed 
temperature
gradients within 700 kpc. 
These components map the overall
temperature
variation from A2029 core to larger
radii similar to other cooling
flow clusters 
\citep{2006PhR...427....1P}.
However, \cite{PaternoMahleretal2013}
found that within
the central 10$\arcsec$
(13 kpc) of A2029, the
gas in the northern
direction is cooler 
(4.5 $\pm$ 0.3 keV)
than in 
the southwest ($\sim$ 5.7 keV).
This asymmetry is
expected, as gas sloshing 
displaces cooler gas from
the cluster core to larger radii
\citep{2023ApJ...944..132S}. 
Our measurements of a
3.42 $\pm$ 0.78
keV component in the central region is 
consistent with 
the cooler gas associated
with the sloshing feature reported
by \cite{PaternoMahleretal2013}.

The indication of the central 2.85 keV gas-component is a significant development. 
The radiative cooling time
of the hot atmosphere drops from $10^9$ yr at an altitude of 20 kpc to about $2\times 10^8~\rm yr$ below 10 kpc. 
Cool gas is observed over a range of temperatures from $\sim 6.8~\rm keV$ through $4~\rm keV$ reaching a floor at 2.8 keV. This temperature drop is spectroscopically consistent with radiative cooling \citep{2006PhR...427....1P}.  Therefore, the 2.85 keV gas is very likely to have cooled to its current temperature.  { Assuming the gas is cooling in pressure balance, the gas density will be a factor of three higher than the $0.15 ~\rm cm^{-3}$ found by Chandra and thus will will occupy a third of the volume.} 
It is expected to { flow inward and become} centrally concentrated with a cooling time 2-3 times below the mean central cooling time of $2\times 10^8~\rm yr$ \citep{2016ApJ...830...79M}. The $\lesssim 10^8~\rm yr$ timescale is low for such a hot cluster. Assuming this temperature structure is long-lived and absent a strong heating mechanism, this reservoir should be cooling to lower temperatures and forming stars. 

Most systems with such short central cooling times have spectacular nebular emission line systems from gas at $10^4$ K \citep{2010ApJ...721.1262M}, $\sim 30$ K molecular gas reservoirs exceeding $10^9 ~\rm M_{\odot}~yr^{-1}$ (Pulido et al. 2018), and  star formation rates of several to several tens of solar masses per year (McDonald et al. 2018).   Yet the 2.85 keV gas reservoir is the coldest gas detected in Abell 2029. The H$\alpha$ luminosity from the central galaxy lies below $<4.4\times 10^{39}~\rm erg~s^{-1}$ \citep{2010ApJ...721.1262M}, and its visible star formation rate lies below $<0.03~\rm M_{\odot}~yr^{-1}$ \citep{2010ApJ...719.1844H}\citep{2012ARA&A..50..455F,2012NJPh...14e5023M} { which appears unable to compensate for cooling losses.}
In these respects, Abell 2029 is a relatively rare outlier in its cooling properties 
\citep{2020ApJ...897...57M,2016ApJ...830...79M}.  XRISM has only deepened the mystery of why it's atmosphere is not cooling into a visible form of matter or perhaps is cooling instead into an invisible form of matter \citep{2023MNRAS.521.1794F}.

\citet{Dullo2017} found that the central galaxy IC1101 has an unusually large 4.2 kpc core signaling a central black hole mass of $4-10\times 10^{10}\rm M_\odot$. 
\citet{Prasad2024}
have suggested that the large black hole mass may promote a steady, hot Bondi mode of black hole accretion that continually staves off cooling, rather than cold, episodic accretion which results in periods of cold gas buildup and star formation. 
Nevertheless, how the jet energy is transported and dissipated into heat remains a mystery. 
With a successful gate-valve opening 
in the future, Resolve will also be 
capable of detecting emission lines 
from O, Mg, Si, and S. 
This will be a big leap forward, 
enabling the detection of much
lower temperatures through 
line-ratio diagnostics.

Additionally, Resolve
detects two cooler gas components
in the N1 region,
with temperatures ranging
4.3 to 4.5 keV. 
As shown in Figure 
\ref{fig:chandra_image}, 
the N1 region is
adjacent to the gas sloshing spiral
contours spatially
resolved by Chandra. 
Cool gas displaced 
from the cluster center
by sloshing can extend beyond
Chandra's detection limits 
due to low surface brightness. 
We speculate that the cooler components 
detected in the N1 region are associated
with sloshed cool gas that
has reached 
the N1 region.
}

\section{Summary}\label{sec:summary}

We present $\sim$500 ks XRISM/Resolve
observations of the A2029 galaxy 
cluster, covering its center out to 
R$_{2500}$. 
This study focuses on investigating 
multi-phase gas using high-resolution
spectra of $\fexxv$ and $\fexxvi$ 
emission lines. 
Below, we summarize our key findings.

\begin{itemize}
\item Due to modest PSF of Resolve, emission 
from a particular region of interest can be 
contaminated by emission from surrounding
regions. 
We account for spectral contamination from surrounding regions by modeling emission from all three regions simultaneously, including the effect of photon leakage between them.
A single-temperature broadband 
fit (2–10 keV) yields best-fit 
temperatures of 6.73 keV
(central), 7.61 keV (N1), 
and 8.14 keV (N2), 
with corresponding metal 
abundances of 0.64, 0.34, 
and 0.41 $Z_{\odot}$. 
A two-temperature fit reveals 
cooler gas components of 
5.14 keV (central) and 3.66 keV 
(N1). 
Velocity broadening is measured
at 157 km/s, $<$82 km/s, 
and 92 km/s in the central,
N1, and N2 regions, respectively.
All results are validated using
both AtomDB and SPEXACT databases.

\item The superb spectral
resolution of Resolve
allows, for the first time, the detection
of several key Fe emission lines in the central,
N1, and N2 regions of A2029. 
We resolve four bright emission lines (w, x, y, z)
in the $\fexxv$ K$\alpha$ line-complex at 6.6--6.7
keV, along with
$\fexxv$ He$\beta$ at 7.882 keV, $\fexxvi$-Ly$\alpha$1 at 6.95, $\fexxvi$-Ly$\alpha$2 at 6.97 keV, 
and $\fexxvi$-Ly$\beta$ at 8.25 keV
in all three regions. 
Additionally, 
we detect the
faint $\fexxv$ He$\gamma$ line
in the central region. 
Detection of these lines enable us to 
measure detailed
multi-temperature gas by using line-ratio
diagnostics.

\item We measure the observed 
line-fluxes of 
 $\fexxv$ and $\fexxvi$ lines
and derive the key line flux ratios,
including
$\fexxv$ $x+y+z/w$, $\fexxv$ He$\beta$/z,
$\fexxv$ He$\gamma$/z, 
$\fexxvi$-Ly$\alpha$/$\fexxv$-He$\beta$,
$\fexxvi$-Ly$\alpha$/$\fexxv$-He$\gamma$,
and $\fexxvi$-Ly$\alpha$/$\fexxv$ z.
By comparing these
observed ratios with predictions
from AtomDB and SPEXACT,
we determine excitation and ionization temperatures
in central, N1, and N2 regions. 
Our analysis 
resolve several cooler gas 
components in the central and
N1 regions, 
significantly lower than 
the best-fit single-temperature
CIE model, indicating the presence
of multi-phase gas.
In contrast, line-ratio-derived temperatures in the N2 region
remain consistent with a
single-temperature CIE model.

\item The line-ratio-derived temperatures, including
5.12, 7.08, 7.59, 8.3 keV 
in the
central region, 8.6 and 9.8 keV 
in the N1 region, 
and $\sim$ 8 keV 
in the N2 region, consistent with the 
temperature gradients within 
700 kpc of A2029
 by previous Chandra and XMM-Newton
observations. These 
temperature components map the
overall temperature variation of 
A2029, consistent with the characteristics of a cool-core galaxy cluster.

\item In the central region,
we identify
two cooler gas components
at 3.42 $\pm$ 0.78 keV
and 2.85$^{+0.75}_{-0.32}$ keV.
The 3.42 $\pm$ 0.78 keV component 
aligns with previously observed displaced cool gas in the northward direction, driven by sloshing within 10$\arcsec$ of the cluster core.
The 2.85 keV component, however, has not been detected in any prior observations.
We suggest that this newly 
detected component may be a direct result of cooling flow within the central 20 kpc of A2029.
Additionally, in the inner northern (N1) region, we detect two cooler gas components at 4.24 keV and 4.55 keV, significantly cooler than best-fit
single-temperature CIE
model.
We argue that these components are associated with cool gas displaced by sloshing motions, reaching the N1 region.
\end{itemize}

{Deeper XRISM observations of the center of A2029 would produce much better constraints on the level of resonant scattering, providing a higher confidence of the coolest temperature component detected by the line ratio method here. They would also reveal lines from Ar, S, and possibly Si, providing more leverage on the cooler temperature components. Should the Resolve Gate Valve open, we will not only have access only to lower-energy emission lines like O, Ne, and Mg, but we will also better constrain the cooler components using a broad-band fit, and directly compare those results to the line ratio method.}

\section*{Acknowledgement}
We thank referee for their insightful comments and suggestions. 
We gratefully acknowledge the dedicated efforts of the many engineers and scientists whose hard work over the years made the XRISM mission possible.
AS and EDM acknowledges support from NASA grants 80NSSC20K0737 and 80NSSC24K0678. 
MS acknowledges support from NASA grant 80NSSC23K0650.
SE acknowledges the financial contribution from the contracts Prin-MUR 2022 supported by Next Generation EU (M4.C2.1.1, n.20227RNLY3 {\it The concordance cosmological model: stress-tests with galaxy clusters}), from the European Union’s Horizon 2020 Programme under the AHEAD2020 project (grant agreement n. 871158), and from the INAF {\it Theory} grant 1.05.24.05.10.
The material is based upon work supported by NASA under award number 80GSFC21M0002.

\bibliographystyle{aasjournal}
\bibliography{sample631} 

\section*{Appendix A: Effect of resonant scattering on optically thick line-flux measurements}\label{appendixA}
{To evaluate the effect of resonant scattering on the measured $\fexxv\ w$ line flux 
at the central region of A2029,
we fitted the Resolve spectra using the {\tt RSAPEC} model \citep{2023ApJ...959..126C}.
Since our primary goal was to estimate the level of systematic uncertainty in the $\fexxv\ w$ line flux,
we restricted the spectral fitting to the 6.6--6.8 keV (rest-frame) energy band.
We fixed $kT$ and abundance to their best-fit values from the 1T model, as reported in Table \ref{tab:best_params}.
Figure \ref{fig:resonance} shows the best-fit model.

\begin{figure}[h!!]
    \includegraphics[width=0.5\textwidth]{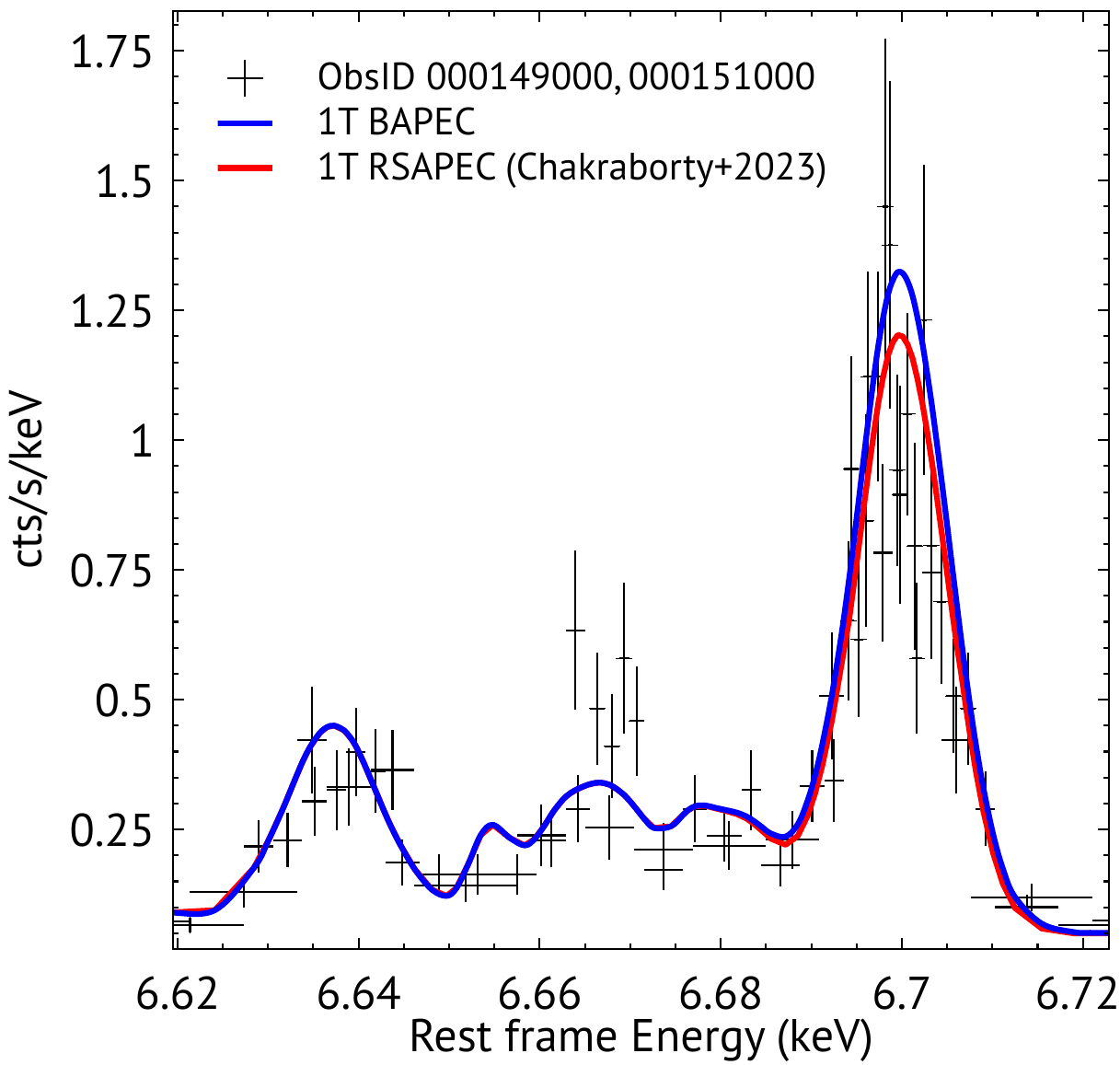}
    \vspace{0.01in}
    \caption{Resolve spectrum extracted from the 
    central region and zoomed into
    $\fexxv$ He$\alpha$ region. Blue and
    red curves show the best-fit {\tt BAPEC}
    and {\tt RSAPEC} model spectra, respectively. {Alt text: Resolve spectrum from the central region is fitted with {\tt BAPEC} and {\tt RSAPEC} models.}
    }
    \label{fig:resonance}
\end{figure}

We note that the 37 ks Resolve exposure is insufficient to directly measure the degree of resonant scattering suppression.
Instead, we derived a 1$\sigma$ 
upper limit of $\sim$12\%, inferred from the best-fit $n_{L}$ parameter
(see \cite{2023ApJ...959..126C} for a detailed discussion of fitting parameters) and comparison with the best-fit 1T {\tt BAPEC} model.
This level of suppression introduces a 1$\sigma$ 
systematic uncertainty of $+50.4$ erg/cm$^{2}$/s
in the intrinsic $\fexxv\ w$ line flux, as reported in Table \ref{tab:gaussian_lines}.
Consequently, this also contributes to a systematic uncertainty in the $\fexxv\ (x + y + z)/w$ line ratio,
and therefore in the inferred temperature.
Figures \ref{fig:line_ratios} and \ref{fig:temp_lineratio} illustrate the resulting updated uncertainties 
(systematic + statistical) in the
measured line ratio and derived temperature, respectively.

It is important to emphasize, however, that the estimated suppression of the $w$-line flux represents a 1$\sigma$ upper limit,
based on the limited 37 ks exposure with Resolve.
}

\section*{Appendix B: Resolve spectrum from the central region fitted with a 3T model }\label{appendixB}
Since we obtained two cooler gas-components
($\sim$ 2.85 keV and 3.42 keV) from the line-ratio diagnostics at the
central region of A2029,
we tried to fit the Resolve
broadband spectrum (2-10 keV) with a 3T
absorbed {\tt BAPEC} model. 
We fixed the two temperature parameters to
2.85 keV and 3.42 keV and tied 
abundance, redshift, and $\sigma_v$
parameters
among
three components. 
Table \ref{tab:3T_table} shows the
best-fit parameters for the 3T spectral 
fitting. 
We note that due to the loss of 
spectral energy band below 2 keV
with the Gate Valve closed, 
directly constraining lower temperature 
gas with a broad-band spectral fit is challenging with Resolve. 
These challenges are likely exacerbated by the relatively short 37-ks exposure on the central region. That we are able to tentatively detect the lower temperature component using line ratios of highly ionized Fe at energies around 6 keV demonstrates the power of this technique.

\begin{table}[h!!]
\caption{Best-fit parameters from the Resolve spectral fitting of the central region using a 3T absorbed {\tt BAPEC} model\label{tab:3T_table}}   
\begin{center}
\begin{tabular}{cccl}
Database & Model & Parameter & Value\\
\hline
\hline
\\
& & $kT_1$ (keV) & 2.85 (fixed)\\
& & $kT_2$ (keV) & 3.42 (fixed)\\
& & $kT_3$ (keV) & 8.40$^{+0.85}_{-0.82}$\\
& & Abundance ($Z_{\odot}$) & 0.69$^{+0.04}_{-0.04}$\\
{\tt AtomDB} & 3T & Redshift & 0.07776$^{+0.00004}_{-0.00004}$\\
& & $\sigma_{v}$ (km/s) & 158$^{+14}_{-13}$\\
& & norm$_1$ (10$^{12}$ cm$^{-5}$) & 1.62 (1$\sigma$ upper-limit)\\
& & norm$_2$ (10$^{12}$ cm$^{-5}$) & 1.90$^{+1.47}_{-1.47}$\\
& & norm$_3$ (10$^{12}$ cm$^{-5}$) & 3.54$^{+0.54}_{-0.54}$\\
& & C-stat/dof & 3855/3852\\
\hline
\end{tabular}
\end{center}
\end{table}

\end{document}